\documentclass[prd, onecolumn]{revtex4}
\usepackage{amsmath}
\usepackage{graphicx}
\usepackage{dcolumn}
\usepackage{bm}
\usepackage{wrapfig}
\usepackage{epsfig}
\usepackage{breqn}
\usepackage{amssymb}
\usepackage{subfigure}
\usepackage[usenames,dvipsnames]{pstricks}
\usepackage{epsfig}
\usepackage{pst-grad} 
\usepackage{pst-plot} 
 \usepackage[usenames,dvipsnames]{pstricks}
 \usepackage{epsfig}
 \usepackage{pst-grad} 
 \usepackage{pst-plot} 
 \usepackage{float}
\usepackage[all]{xy}

\begin{document}

\preprint{APS/123-QED}

\title{On The Dynamics of a Closed Viscous Universe}
\author{Ikjyot Singh Kohli}
\email{isk@yorku.ca}
\affiliation{York University - Department of Physics and Astronomy}
\author{Michael C. Haslam}
\email{mchaslam@mathstat.yorku.ca}
\affiliation{
York University - Department of Mathematics and Statistics
}
\date{November 1, 2013}

\begin{abstract}
We use a dynamical systems approach based on the method of orthonormal frames to study the dynamics of a non-tilted Bianchi Type IX cosmological model with a bulk and shear viscous fluid source. We begin by completing a detailed fix-point analysis which give the local sinks, sources and saddles of the dynamical system. We then analyze the global dynamics by finding the $\alpha$-and $\omega$-limit sets which give an idea of the past and future asymptotic behavior of the system. The fixed points were found to be a flat Friedmann-LeMa\^{i}tre-Robertson-Walker (FLRW) solution, Bianchi Type $II$ solution, Kasner circle, Jacobs disc, Bianchi Type $VII_{0}$ solutions, and several closed FLRW solutions in addition to the Einstein static universe solution. Each equilibrium point was described in both its expanding and contracting epochs. We conclude the paper with some numerical experiments that shed light on the global dynamics of the system along with its heteroclinic orbits. With respect to past asymptotic states, we were able to conclude that the Jacobs disc in the expanding epoch was a source of the system along with the flat FLRW solution in a contracting epoch. With respect to future asymptotic states, we were able to show that the flat FLRW solution in an expanding epoch along with the Jacobs disc in the contracting epoch were sinks of the system. We were also able to demonstrate a new result with respect to the Einstein static universe. Namely, we gave certain conditions on the parameter space such that the Einstein static universe has an associated stable subspace. We were however, not able to conclusively say anything about whether a closed FLRW model could be a past or future asymptotic state of the model.

\end{abstract}
\maketitle 


\section{Introduction}
In this paper, we analyze a non-tilted Bianchi Type IX cosmological model with a viscous fluid source containing constant bulk and shear viscous coefficients, while neglecting the effects of heat conduction. Such a cosmological model is homogeneous on three-dimensional spacelike orbits of the isometry group $G_{3}$, and is therefore spatially homogeneous (Page 22-23, \cite{ellis}). We take the dimension of the isotropy subgroup to be $d=0$, hence considering a spatially homogeneous and anisotropic model of the universe. In this case, the Einstein field equations reduce to a coupled system of nonlinear first-order ordinary autonomous differential equations. One can then use methods of analyzing dynamical systems to obtain important information about the dynamical evolution of such a universe model, with particular emphasis on past and future asymptotic states. 

The Bianchi Type IX model is perhaps among the most well-known and well-studied models in cosmology. Belinskii and Khalatnikov \cite{belinksii1}  investigated the evolutionary dynamics of the Bianchi Type IX metric as it approached an initial singularity. Belinksii, Lifshitz, and Khalatnikov \cite{lifshitz} showed that near the initial singularity, the Bianchi IX model exhibits oscillatory behavior represented by a series of Kasner-like ``bounces''. Misner  \cite{misnerb} \cite{misner} applied Hamiltonian methods to show that the dynamics of a Bianchi IX model is equivalent to the classical problem of a particle in a potential well. In the former paper, he formulated a quantum theory based on this geometry by setting up canonical commutation relations on the independent canonical variables. In the latter paper, Misner introduced the well-known Mixmaster universe as an attempt to describe the present-day spatially homogeneous and isotropic universe as a result of a ``smoothing-out'' process of early-universe anisotropy. Matzner, Shepley, and Warren \cite{matzner1} performed a detailed analytical and numerical analysis of Bianchi Type IX models containing dust. They were able to prove the existence of regions of infinite density and a time of maximum expansion. Ryan \cite{ryan1} \cite{ryan2} \cite{ryan3} using the Hamiltonian methods of Arnowitt, Deser, and Misner \cite{admnew} analyzed Bianchi Type IX universes which simultaneously exhibited expansion, rotation, and shear, and placed particular emphasis on the dynamics near the initial singularity in his analysis. Ryan also showed that the dynamical equations simplify in the region near the initial singularity, and used the concept of a point moving in a set of potentials to analyze the dynamical behavior near the singularity. Barrow and Matzner \cite{barrowmatz} analyzed the evolution of a massive scalar field in Bianchi Type IX model. They were able to show that the probability of a bouncing epoch occurring at very early times is infinitesimally small.

The problem of recollapse is one of the central themes of this paper, and we will revisit it in the next section, when deciding how to normalize our dynamical variables to allow for the possibility of our model expanding or contracting. Barrow and Tipler \cite{barrowtip1} first showed that the existence of a maximal hypersurface is a necessary and sufficient condition for the existence of a final singularity in a universe with a compact Cauchy surface. They further showed that a cosmological model with topology $S^3$ can admit such maximal hypersurfaces. Barrow, Galloway and Tipler \cite{barrowtip2} formulated the closed-universe recollapse conjecture, where they showed that if the positive pressure criterion, dominant energy and matter regularity conditions hold, then a FLRW universe with topology $S^3$ must recollapse. They also considered a number of Bianchi Type IX universes with various matter tensors, and provided a new recollapse conjecture for such matter-filled universes. Barrow \cite{barrownuc1} discussed in detail the question of whether closed universes can avoid recollapsing before an inflationary period ensues. It was shown that closed universes in an extreme initial anistropic state cannot recollapse until they are close to isotropy. Barrow made the point that even if a universe possesses a $S^3$ topology and the strong energy condition holds, it is not known whether all anisotropic closed universes recollapse. Related to this, Calogero and Heinzle \cite{caloheinzle} proved that there exists a class of Bianchi Type IX models that obey the strong energy condition but do not recollapse, rather, they expand for all times. Lin and Wald \cite{linwald} \cite{linwald2} showed that for matter which satisfies the dominant energy condition in addition to having nonnegative average principal pressures, there is no corresponding Bianchi Type IX model that expands for an infinite time. Wald \cite{waldix} examined the future asymptotic behavior of initially expanding spatially homogeneous models containing a positive cosmological constant.  It was shown that if the cosmological constant, $\Lambda$ is sufficiently large compared with spatial-curvature terms, the Bianchi Type IX model exhibits stable future asymptotic behavior only in the case of recollapse and an asymptotic late-time approach to a de Sitter spacetime. 

Burd, Buric, and Ellis \cite{burdburicellis} performed a detailed study of the chaotic behavior of the Bianchi Type IX model. They numerically calculated the Lyapunov exponent and showed that it decreases steadily. In contradiction to this result, Rugh and Jones \cite{rughjones} showed that the maximal Lyapunov exponent for the phase flow is zero with respect to the time variable used in previous studies for the Bianchi IX model, but concluded that the deterministic model is unpredictable due to a large nonnegative entropy. Uggla and Zur-Muhlen \cite{ugglazur} investigated locally rotationally symmetric Bianchi Type IX models with a perfect fluid source. By considering a locally rotationally symmetric model, they obtained a reduced first-order system of differential equations that allowed them so see the full set of solutions from the initial big bang singularity to the final big crunch singularity.  Cornish and Levin \cite{cornishlevin} \cite{cornishlevin2} used coordinate-invariant fractal methods to show that the Bianchi Type IX model is indeed chaotic, independent of any analysis using the methods of calculating Lyapunov exponents.
Rendall \cite{rendall} showed that for Bianchi IX models whose observers have worldlines orthogonal to the spatial hypersurfaces, no singularity can occur in finite time. Rendall \cite{rendall2} also performed a detailed analysis of the asymptotic behavior of the Bianchi Type IX model. It was shown that there are infinitely many oscillations near the singularity and that the Kretschmann scalar is unbounded in that region. Van Den Hoogen and Olasagasti \cite{vandenolas}
investigated the isotropization of the Bianchi Type IX model with an exponential potential field. They found conditions on the potential exponent that classified inflationary and isotropization behavior. Ringstr\"{o}m \cite{ringstrom1} used dynamical system methods to investigate the asymptotic behavior of the Bianchi Type IX model with an orthogonal perfect fluid close to the initial singularity. It was found that in the case of a stiff fluid, the solution converges to a point. For other types of matter, the solutions converge to an attractor consisting of Bianchi Type II vacuum orbits. De Oliveira et.al. \cite{deoliv1} analyzed the dynamics of a Bianchi Type IX model with comoving dust and a cosmological constant, in which they found evidence of homoclinic chaos in the dynamical evolution. Barrow, Ellis, Maartens, and Tsagas \cite{barrowellismaartenstsagas} showed that spatially homogeneous Bianchi Type IX models destabilize an Einstein static universe. Heinzle, R\"{o}hr, and Uggla \cite{heinzleuggla} investigated a locally rotationally symmetric Bianchi Type IX model with an orthogonal perfect fluid source. They showed that when the perfect fluid obeys the strong energy condition, such a model expands from an initial singularity and recollapses to a singularity. Heinzle and Uggla \cite{heinzleuggla2} considered the past asymptotic dynamics of both a Bianchi Type IX vacuum and an orthogonal perfect fluid model. They formulated precise conjectures regarding the past asymptotic states using a dynamical systems approach. They also made detailed comparisons with previous metric and Hamiltonian approaches that also analyzed dynamical behavior in the neighborhood of the initial singularity. Heinzle and Uggla \cite{heinzleuggla3} also gave a new proof of the Bianchi type IX attractor theorem which states that the past-time asymptotic behavior of the Bianchi Type IX solutions is determined  by Bianchi Type I and II vacuum states.  Calogero and Heinzle \cite{caloheinzle2} considered among other models a locally rotationally symmetric Bianchi Type IX model containing Vlasov/collisionless matter, elastic matter, and magnetic fields. They discovered that generic type IX solutions oscillate toward the initial singularity. Barrow and Yamamoto \cite{barrowyama} studied the stability of the Einstein static universe as a non-locally rotationally symmetric, that is, a general Bianchi Type IX model with both non-tilted and tilted perfect fluids. They showed that the Einstein static universe is unstable to homogeneous perturbations of the Bianchi Type IX model to both the future and the past. Uggla \cite{ugglanew} has described recent developments with respect to oscillatory spacelike singularities in Bianchi Type IX models.


As discussed by Gr{\o}n and Hervik (Chapter 13, \cite{hervik}), viscous models have become of general interest in early-universe cosmologies largely in two contexts.  Firstly, in models where bulk viscous terms dominate over shear terms, the universe expands to a de Sitter-like state, which is a spatially flat universe neglecting ordinary matter, and including only a cosmological constant. Such models isotropize indirectly through the massive expansion. Secondly, in the absence of any significant heat flux, shear viscosity is found to play an important role in models of the universe at its early stages. In particular, neutrino viscosity is considered to be one of the most important factors in the isotropization of our universe. 
Parnovskii \cite{parnovskii} for example, investigated the influence of viscosity on the dynamics of a Bianchi type II universe, where it was shown that at late times such a universe approaches an FLRW type singularity or an anisotropic solution. It was further shown that the bulk viscosity has an important influence in creating entropy per particle throughout the future evolution of such a model. Misner \cite{misnervisc1} considered solutions of the Einstein field equations with flat homogeneous spacelike hypersurfaces with anisotropic expansion rates in which effects of viscosity were included  in the associated radiation. Misner \cite{misnervisc2} also studied the effect of neutrino viscosity on the homogeneous anisotropy in relation to the expansion of the early universe. 


Bianchi Type IX models containing viscous fluids have also been studied in some detail. Caderni and Fabbri \cite{cadfab} \cite{cadfab2} investigated the isotropization of the Bianchi Type IX model due to neutrino viscosity. Banerjee and Santos \cite{banerjeesantos} studied the dynamical effects of viscous fluids on the Bianchi Type IX model. Banerjee, Sanyal, and Chakraborty \cite{banerjee2} found exact solutions to the Einstein field equations for a Bianchi Type IX model with a viscous fluid distribution. Chakraborty and Chakraborty \cite{chakraborty} investigated the dynamics of a Bianchi Type IX model with a bulk viscous fluid and variable gravitational and cosmological constants. Pradhan, Srivastav, and Yadav \cite{prasriya} examined the dynamics of a Bianchi Type IX model with a varying cosmological constant and both bulk and shear viscosities. Bali and Yadav \cite{baliyadav} investigated a Bianchi Type IX model with a viscous fluid containing both bulk and shear viscosities. 

All of the aforementioned methods that consider viscous fluids employ the metric formalism of general relativity and assume supplemental conditions between the different metric components in order to obtain exact solutions. In this paper, we will use dynamical systems methods built upon the pioneering framework of orthonormal frames initiated by Ellis and MacCallum \cite{ellismac} to analyze the behavior of the Bianchi Type IX model with a viscous fluid with respect to early times, late times, and intermediate times. Dynamical systems methods have been used to study viscous cosmologies by van den Hoogen and Coley \cite{vdh}, and Kohli and Haslam \cite{isk1} \cite{isk2}. As we mentioned above, and explicitly state below, in our model, the bulk and shear viscosity coefficients are taken to be nonnegative constants. The case where these coefficients are not constant has been studied in some detail. In particular, Barrow \cite{barrownuc2} showed that models of an inflationary universe driven by Witten strings in the very early universe are equivalent to the addition of bulk viscosity to perfect fluid cosmological models with zero curvature. In this work, Barrow considered the case where the bulk viscosity has a power-law dependence upon the matter density. It was shown that if the exponent is greater than $1/2$, there exist deflationary solutions which begin in a de Sitter state and evolve away from it asymptotically in the future. On the other hand, if this exponent is less than $1/2$ (which includes the case considered in our present work), then solutions expand from an initial singularity towards a de Sitter state. Barrow \cite{barrownuc3} estimated the entropy production associated with anisotropy damping in the early universe by considering a Bianchi type I metric with an equilibrium radiation gas and anisotropic stresses produced by shear viscosity. It was shown that the shear viscosity based on kinetic theory has the general form of being proportional to the matter density and that the entropy production due to collisional transport is negligible in such a model.

Our approach will allow one to fully ascertain the effects of the constant bulk and shear viscous coefficients on the dynamics of the Bianchi Type IX model, and will therefore be more general than the metric approaches taken so far. 
We note that to the best of the authors' knowledge at the time of writing this article, such an approach based on dynamical systems theory has not been investigated in the literature. 

\section{The Viscous Fluid Matter Source}
In the absence of heat conduction, the energy-momentum tensor corresponding to a viscous fluid with fluid velocity four-vector is given by \cite{isk1}
\begin{equation}
\label{eq:enmom}
T_{ab} = \left(\mu + p\right)u_{a}u_{b} + g_{ab}p - 3\xi H h_{ab} - 2\eta \sigma_{ab},
\end{equation}
where $\mu$, $p$, and $\sigma_{ab}$ denote the fluid's energy density, pressure, and shear respectively, while $\xi$ and $\eta$ denote the bulk and shear viscosity coefficients of the fluid. Throughout this paper, both coefficients are taken to be \emph{nonnegative constants}. $H$ denotes the Hubble parameter, and $h_{ab} \equiv u_{a} u_{b} + g_{ab}$ is the standard projection tensor corresponding to the metic signature $\left(-,+,+,+\right)$.

We additionally assume that this fluid obeys a barotropic equation of state, $p = w \mu$, where $w:\{w \in \mathbb{R}: -1 \leq w \leq 1\}$, is an equation of state parameter. Some typical values for $w$ are $w = 0$ (dust), $w = -1$ (cosmological constant), $w = 1/3$ (radiation), and $w = 1$ (stiff fluid). In order to derive a set of evolution equations for the model, we will need expressions for the total energy density, $\bar{\mu}$, total pressure $\bar{p}$, and total anisotropic stress $\bar{\pi}_{ab}$. Using the definitions,
\begin{equation}
\label{totalquants}
\bar{\mu} = T_{ab} u^{a} u_{b}, \quad \bar{p} = \frac{1}{3}h^{ab}T_{ab}, \quad \bar{\pi}_{ab} = h_{a}^{c} h_{b}^{d} T_{cd} - \bar{p}h_{ab},
\end{equation}
we find that
\begin{equation}
\label{totalquants2}
\bar{\mu} = \mu, \quad \bar{p} = w \mu, \quad \mbox{and} \quad \bar{\pi}_{ab} = -2 \eta \sigma_{ab}.
\end{equation}

\section{The Dynamical Equations}
The Bianchi cosmologies are described by a four-dimensional pseudo-Riemannian manifold $\mathcal{M}$, a corresponding metric tensor $\mathbf{g}$ defined on $\mathcal{M}$, and a fundamental four-velocity $\mathbf{u}$ that we will take to be orthogonal to the group orbits.  Denoting the orthonormal basis vectors by $\mathbf{e}_{\alpha}$ and the unit vector normal to the orbits of $G_{3}$ by $\mathbf{n}$, and using the quantities in Eq. \eqref{totalquants2}, the Einstein field equations take the form (Page 39, \cite{ellis}):
\begin{eqnarray}
\label{efe1}
\dot{H} &=& -H^2 - \frac{2}{3}\sigma^2 - \mu \left(\frac{1}{6} + \frac{1}{2}w\right), \\
\label{efe2}
\dot{\sigma}_{ab} &=& -3H \sigma_{ab} + 2 \epsilon^{uv}_{(a}\sigma_{b)u}\Omega_{v} - S_{ab} - 2\eta \sigma_{ab}, \\
\label{efe3}
\mu &=& 3H^2 - \sigma^2 + \frac{1}{2} R, \\
\label{efe4}
0 &=& 3 \sigma^{u}_{a} a_{u} - \epsilon_{a}^{uv} \sigma_{u}^{b} n_{bv},
\end{eqnarray}
where $S_{ab}$ and $R$ are the three-dimensional spatial curvature and Ricci scalar and are defined as:
\begin{eqnarray}
\label{scalars1}
S_{ab} &=& b_{ab} - \frac{1}{3}b^{u}_{u}\delta_{ab} - 2 \epsilon^{uv}_{(a}n_{b)u}a_{v}, \\
\label{scalars2}
R &=& -\frac{1}{2}b^{u}_{u} - 6 a_{u} a^{u},
\end{eqnarray}
where $b_{ab} = 2 n_{a}^{u} n_{ub} - \left(n^{u}_{u}\right) n_{ab}$. We have also denoted by $\Omega_{v}$ the angular velocity of the spatial frame.
The matrix $n_{ab}$ and vector $a_{c}$ are used in decomposing the structure constants of $G_{3}$ and classify the Bianchi cosmologies (See Page 36, \cite{ellis} for more details). Using the Jacobi identities, one obtains evolution equations for these variables as well:
\begin{eqnarray}
\label{efe5}
\dot{n}_{ab} &=& -H n_{ab} + 2 \sigma_{(a}^u n_{b)u} + 2 \epsilon^{uv}_{(a}n_{b)u}\Omega_{v}, \\
\label{efe6}
\dot{a}_{a} &=& -H a_{a} \sigma_{a}^{b} a_{b} + \epsilon_{a}^{uv} a_{u} \Omega_{v}, \\
\label{efe7}
0 &=& n_{a}^{b} a_{b}.
\end{eqnarray}
The contracted Bianchi identities give the evolution equation for $\mu$ as (Page 40, \cite{ellis})
\begin{equation}
\label{mudot}
\dot{\mu} = -3H\left(\mu + p\right) - \sigma_{a}^{b} \pi_{b}^{a} + 2a_{a}q^{a}.
\end{equation}

The algebraic constraints for the Bianchi Type IX model are
\begin{equation}
\label{typeixcons}
a_{a} = 0, \quad n_{ab} = \mbox{diag} \left(n_{11}, n_{22}, n_{33}\right), \mbox{ where } n_{11} > 0, \quad n_{22} > 0, \quad n_{33} > 0.
\end{equation}
Note that, for the Bianchi class A models, it can also be shown that $\Omega_{v} = 0$.

The standard way to proceed from this point is to use expansion-normalized variables, for which one reduces the dimension of the state space by introducing a dimensionless time variable $\tau$. In this approach, the Raychaudhuri equation Eq. \eqref{efe1} decouples from the system of differential equations, yielding a reduced system of autonomous first-order ordinary differential equations. The problem with using this method for the Bianchi Type IX model is that as we discussed in the introduction, the Bianchi Type IX model has the potential to recollapse. The notion of recollapse has been investigated for cosmological models whose spatial sections have topology $S^{3}$ or $S^{2} \times S^{1}$ (see the references in the introduction of this article). It should be noted that such models do not always recollapse, but have the potential to do so. Therefore, in order to get a complete picture of the dynamics of the system, one needs to employ a different normalization than the standard expansion-normalized variables. 

We will make use of the approach outlined by Hewitt, Uggla, and Wainwright (Chapter 8, \cite{ellis}) in which $H$ assumes all real values so as to include a re-collapsing epoch $(H < 0)$.  The evolving state vector has the form $\mathbf{x} = \left(\sigma_{+}, \sigma_{-}, n_{1}, n_{2}, n_{3}\right)$, where we have defined:
\begin{equation}
\sigma_{+} \equiv \frac{1}{2}\left(\sigma_{22} + \sigma_{33}\right), \quad \sigma_{-} \equiv \frac{1}{2\sqrt{3}} \left(\sigma_{22} - \sigma_{33}\right),
\end{equation}
and
\begin{equation}
n_{11} \equiv n_{1}, \quad n_{22} \equiv n_{2}, \quad n_{33} \equiv n_{3}.
\end{equation}
We will normalize this state vector by a normalization factor $D$, defined by
\begin{equation}
\label{Ddef}
D \equiv \sqrt{H^2 + \frac{1}{4} \left(n_{1} n_{2} n_{3}\right)^{2/3}}. 
\end{equation}
The resulting state vector is given by
\begin{equation}
\mathbf{\tilde{x}} = \left(\tilde{H}, \tilde{\Sigma}_{+}, \tilde{\Sigma}_{-}, \tilde{N}_{1}, \tilde{N}_{2}, \tilde{N}_{3}\right),
\end{equation}
where
\begin{equation}
\label{huwvars}
\tilde{H} = \frac{H}{D}, \quad \tilde{\Sigma}_{\pm} = \frac{\sigma_{\pm}}{D}, \quad \tilde{N}_{\alpha} = \frac{n_{\alpha}}{D}.
\end{equation}
These variables satisfy the constraint
\begin{equation}
\label{constr1}
\tilde{H}^{2} + \frac{1}{4} \left(\tilde{N}_{1} \tilde{N}_{2} \tilde{N}_{3} \right)^{2/3} = 1.
\end{equation}
We will additionally define a dimensionless time variable $\tilde{\tau}$ such that
\begin{equation}
\label{taudef1}
\frac{dt}{d\tilde{\tau}} = \frac{1}{D}.
\end{equation}
Hewitt, Uggla, and Wainwright then obtain the evolution equation for $D$ as
\begin{equation}
\label{devolve}
\frac{dD}{d \tilde{\tau}} = -\left(1 + \tilde{q}\right)\tilde{H} D,
\end{equation}
where
\begin{equation}
\tilde{q} = \tilde{H}^{2} q.
\end{equation}
We will also define several quantities in addition to Eq. \eqref{huwvars} that will be needed in deriving the full set of evolution equations and their corresponding constraints. Analogous to the case of expansion-normalized variables as found in the Appendix of \cite{hewittbridsonwainwright}, we have
\begin{equation}
\label{huwvars2}
\tilde{S}_{ij} = \frac{R_{\langle{ij}\rangle}}{D^2}, \quad \tilde{\Omega} = \frac{\bar{\mu}}{3D^2}, \quad \tilde{P} = \frac{\bar{p}}{3D^2}, \quad \tilde{\Pi}_{ij} = \frac{\bar{\pi}_{ij}}{D^2},  \quad 3\tilde{\xi}_{0} = \frac{\xi}{D}, \quad 3\tilde{\eta}_{0} = \frac{\eta}{D}, \quad \hat{\Sigma}^2 = \frac{\sigma^2}{3D^2},
\end{equation}
where expressions for $\bar{\mu}, \bar{p}$, and $\bar{\pi}_{ij}$ were derived in Eq. \eqref{totalquants2}. Note that we have additionally used the abbreviation $\hat{\Sigma}^2 = \tilde{\Sigma}_{+}^2 + \tilde{\Sigma}_{-}^2$. The angled brackets indicate that the projected symmetric trace-free components are to be taken. Note that for clarity in notation, we have left off the standard ``three'' superscript on $R_{\langle{ij}\rangle}$. In this paper, it is to be assumed that $R_{ij}$ indicates the three-dimensional Ricci curvature, and $R$ the corresponding three-dimensional Ricci scalar.

First applying the Bianchi Type IX algebraic constraints as given in Eq. \eqref{typeixcons} to Eqs. \eqref{efe1}, \eqref{efe2}, \eqref{efe6}, and \eqref{mudot}, and then normalizing these equations according to Eqs. \eqref{Ddef}, \eqref{huwvars}, \eqref{huwvars2}, \eqref{taudef1} and \eqref{devolve} we obtain the full set of evolution equations as
\begin{eqnarray}
\label{syseqs1}
\tilde{H}' &=& -(1-\tilde{H}^2) \tilde{q}, \\
\label{syseqs2}
\tilde{\Sigma}_{+}' &=& \tilde{\Sigma}_{+} \tilde{H}\left(-2+\tilde{q}\right) - 6 \tilde{\Sigma}_{+} \tilde{\eta}_{0} - \tilde{S}_{+}, \\
\label{syseqs3}
\tilde{\Sigma}_{-}' &=& \tilde{\Sigma}_{-} \tilde{H} \left(-2 + \tilde{q}\right) - 6 \tilde{\Sigma}_{-} \tilde{\eta}_{0} - \tilde{S}_{-}, \\
\label{syseqs4}
\tilde{N}_{1}' &=& \tilde{N}_{1} \left(\tilde{H} \tilde{q} - 4\tilde{\Sigma}_{+}\right), \\
\label{syseqs5}
\tilde{N}_{2}' &=& \tilde{N}_{2} \left(\tilde{H} \tilde{q} + 2\tilde{\Sigma}_{+} + 2 \sqrt{3}\tilde{\Sigma}_{-}\right), \\
\label{syseqs6}
\tilde{N}_{3}' &=& \tilde{N}_{3} \left(\tilde{H} \tilde{q} + 2\tilde{\Sigma}_{+} - 2\sqrt{3} \tilde{\Sigma}_{-}\right), \\
\label{syseqs7}
\tilde{\Omega}' &=& \tilde{\Omega} \tilde{H}  \left(-1 + 2\tilde{q} - 3w\right) + 9 \tilde{H}^{2} \tilde{\xi}_{0} + 12 \tilde{\eta}_{0} \left(\tilde{\Sigma}_{+}^2 + \tilde{\Sigma}_{-}^2\right),
\end{eqnarray}
where
\begin{equation}
\label{qdef}
\tilde{q} = 2\left(\tilde{\Sigma}_{+}^2 + \tilde{\Sigma}_{-}^2\right) + \frac{1}{2}\tilde{\Omega}\left(1+3w\right) - \frac{9}{2}\tilde{\xi}_{0}\tilde{H}.
\end{equation}
The variables $\tilde{S}_{\pm}$ were obtained by normalizing the components of the trace-free spatial Ricci tensor given by Eq. (6.6) in \cite{ellis}, and computed to be
\begin{eqnarray}
\label{Svalues}
\tilde{S}_{+} &=& \frac{1}{6} \left[ \left(\tilde{N}_{2} - \tilde{N}_{3} \right)^2 - \tilde{N}_{1} \left(2\tilde{N}_{1} - \tilde{N}_{2} - \tilde{N}_{3}\right)\right], \\
\tilde{S}_{-} &=& \frac{1}{2\sqrt{3}} \left[ \left(\tilde{N}_{3} - \tilde{N}_{2} \right) \left(\tilde{N}_{1} - \tilde{N}_{2} - \tilde{N}_{3} \right) \right].
\end{eqnarray}
The additional constraint on the dynamical system Eqs. \eqref{syseqs1}-\eqref{syseqs7} is given by the generalized Friedmann equation, \eqref{efe3}. We first note that the Ricci scalar as defined in Eq. \eqref{scalars2} upon applying the Bianchi Type IX algebraic constraints Eq. \eqref{typeixcons} takes the form  \begin{equation}
\label{ricci1}
R = -\frac{1}{2}\left[n_{1}^2 + n_{2}^2 + n_{3}^2 - 2\left(n_{1}n_{2} + n_{2}n_{3} + n_{3}n_{1}\right)  \right].
\end{equation}
Applying Eqs. \eqref{huwvars}, \eqref{huwvars2}, and \eqref{ricci1} to Eq. \eqref{efe3}, we obtain
\begin{equation}
\label{omegadef2}
\tilde{\Omega} = \tilde{H}^2 - \hat{\Sigma}^2 + \frac{R}{6D^2}.
\end{equation}
Upon further applying the constraint Eq. \eqref{constr1} to Eq. \eqref{omegadef2}, we obtain
\begin{equation}
\label{modfriedmann}
\tilde{\Omega} + \hat{\Sigma}^2 + \tilde{V} = 1,
\end{equation}
where
\begin{equation}
\label{Vdef}
\tilde{V} = \frac{1}{12} \left[\tilde{N}_{1}^2 + \tilde{N}_{2}^2 + \tilde{N}_{3}^2 - 2 \tilde{N}_{1} \tilde{N}_{2} - 2\tilde{N}_{2}\tilde{N}_{3} - 2\tilde{N}_{3}\tilde{N}_{1} + 3 \left(\tilde{N}_{1} \tilde{N}_{2} \tilde{N}_{3} \right)^{2/3} \right].
\end{equation}
An important point to note is that from Eq. \eqref{constr1} we have that
\begin{equation}
\label{Hrange}
-1 \leq \tilde{H} \leq 1.
\end{equation}
We also can see from Eq. \eqref{Vdef} that
\begin{equation}
\tilde{V} \geq 0.
\end{equation}
These subsequent conditions were also noted on Page 181 in \cite{ellis}.

Before we proceed, we feel that a technical point is in order. In the standard expansion-normalized variables approach to analyzing the Bianchi cosmologies, one typically reduces the dimension of the dynamical system state space by using the generalized Friedmann equation to eliminate $\tilde{\Omega}$ from the system of equations, thereby making $\tilde{\Omega}'$ an auxiliary equation. The problem with this is that in our approach because the constraint equation for $\tilde{\Omega}$ Eq. \eqref{modfriedmann} contains $\tilde{V}$, the Jacobian matrix will not be defined at all equilibrium points. For example, if were to use Eq. \eqref{modfriedmann} to eliminate $\tilde{\Omega}$ from the system of equations, each term would be replaced with a term that contained $\tilde{V}$ as given by Eq. \eqref{Vdef}, in particular, a factor of $\left(\tilde{N}_{1} \tilde{N}_{2} \tilde{N}_{3}\right)^{-1/3}$ would enter into each equation, which of course is not defined when any one of $\tilde{N}_{1,2,3} = 0$, even if both constraint equations \eqref{constr1} \eqref{modfriedmann} above are satisfied. Therefore, we will not in this paper eliminate $\tilde{\Omega}$ from the dynamical system of equations, as we wish to ascertain the dynamical behavior of all possible equilibrium points. This same methodology was employed by Barrow and Yamamoto \cite{barrowyama}.

\section{A Qualitative Analysis of the Dynamical System}
With the dynamical equations \eqref{syseqs1}-\eqref{syseqs7} and their constraints \eqref{constr1} and \eqref{modfriedmann} in hand, we are in position to perform a detailed analysis of the fixed points of the system. Before we proceed, however, we would like to note some important qualitative properties that can be deduced from the dynamical system. 

\subsection{Symmetries and Invariant Sets}
We note that the dynamical system given by Eqs. \eqref{syseqs1}-\eqref{syseqs7} has three symmetries given by
\begin{eqnarray}
\left[\tilde{H}, \tilde{\Sigma}_{+}, \tilde{\Sigma}_{-}, \tilde{N}_{1}, \tilde{N}_{2}, \tilde{N}_{3}, \tilde{\Omega}\right] &\rightarrow& \left[\tilde{H}, \tilde{\Sigma}_{+}, \tilde{\Sigma}_{-}, -\tilde{N}_{1}, \tilde{N}_{2}, \tilde{N}_{3}, \tilde{\Omega}\right], \\
\left[\tilde{H}, \tilde{\Sigma}_{+}, \tilde{\Sigma}_{-}, \tilde{N}_{1}, \tilde{N}_{2}, \tilde{N}_{3}, \tilde{\Omega}\right] &\rightarrow& \left[\tilde{H}, \tilde{\Sigma}_{+}, \tilde{\Sigma}_{-}, \tilde{N}_{1}, -\tilde{N}_{2}, \tilde{N}_{3}, \tilde{\Omega}\right], \\
\left[\tilde{H}, \tilde{\Sigma}_{+}, \tilde{\Sigma}_{-}, \tilde{N}_{1}, \tilde{N}_{2}, \tilde{N}_{3}, \tilde{\Omega}\right]  &\rightarrow& \left[\tilde{H}, \tilde{\Sigma}_{+}, \tilde{\Sigma}_{-}, \tilde{N}_{1}, \tilde{N}_{2}, -\tilde{N}_{3}, \tilde{\Omega}\right].
\end{eqnarray}
The dynamical system is therefore invariant with respect to spatial inversions in the functions $\tilde{N}_{1}, \tilde{N}_{2}$, and $\tilde{N}_{3}$, which implies that we can take $\tilde{N}_{1} \geq 0$, $\tilde{N}_{2} \geq 0$, and $\tilde{N}_{3} \geq 0$. We can also see that these correspond to the invariant sets of the system. Recall that if we let $M$ be phase space of the flow of the dynamical system, then an invariant set is a set $A \subset M$ such that $g^{t} A = A$, $\forall$ $t$, where $\{g^{t}\}$ represents the  dynamical system on the phase space $M$, and $t \in \mathbb{R}$. In other words, the invariant set consists of entire trajectories \cite{arnolddyn}. Tavakol (Chapter 4, \cite{ellis}) discusses a simple way to obtain the invariant sets of a dynamical system. Let us consider a dynamical system $\dot{x} = v(x), \quad x \in \mathbb{R}^{7}$. Let $Z: \mathbb{R}^{7} \to \mathbb{R}$ be a $C^{1}$ function such that $Z' = \alpha Z$, where $\alpha: \mathbb{R}^{7}$ is a continuous function. Then the subsets of $\mathbb{R}^{7}$ defined by $Z > 0$, $Z = 0$, and $Z < 0$ are invariant sets of the flow of the dynamical system. Applying this proposition to our dynamical system in combination with the symmetries found above, we see that $\tilde{N}_{i} > 0$ and $\tilde{N}_{i} = 0$, where $i = 1,2,3$ are invariant sets of the system. 

Combinations of $\tilde{N}_{i} > 0$ and $\tilde{N} = 0$ determine various Bianchi types of Class A. However, because of the constraint \eqref{constr1}, these combinations necessarily restrict the value of $\tilde{H}$ as well. We list these Bianchi invariant sets based on the description given on Page 126 of \cite{ellis} in Table ~\ref{table1}.
\begin{table}[H]
\caption{The various Bianchi invariant sets, with $\alpha = 1,2,3$.}
\begin{center}
\begin{tabular}{|c|c|c|}
\hline
Notation & Restrictions on $\tilde{N}_{i} \geq 0$ & Restriction on $\tilde{H}$  \\ \hline
$B(I)$ & All zero & $\tilde{H}^2 = 1$  \\ \hline
$B_{\alpha}(II)$ & One non-zero & $\tilde{H}^2 = 1$  \\ \hline
$B_{\alpha}(VII_{0})$ & Two non-zero & $\tilde{H}^2 = 1$  \\ \hline
$B(IX)$ & All non-zero & $\tilde{H}^2 + \frac{1}{4} \left(\tilde{N}_{1} \tilde{N}_{2} \tilde{N}_{3}\right)^{2/3} = 1$ \\ \hline
\end{tabular}
\end{center}
\label{table1}
\end{table}

The dynamical system also admits \emph{shear invariant sets} which arise from enforcing certain restrictions on the shear variables. It follows from Eqs. \eqref{syseqs1}-\eqref{syseqs7} that
\begin{eqnarray}
\label{shearsets}
\tilde{\Sigma}_{-} = 0 &\Rightarrow& \tilde{N}_{2} = \tilde{N}_{3} > 0, \quad \tilde{N}_{1} = 0, \quad \tilde{H} = \pm 1, \quad \mbox{for Bianchi Types $VII_{0}$, $IX$}, \\
\tilde{\Sigma}_{-} = 0 &\Rightarrow& \tilde{N}_{2} = \tilde{N}_{3} = 0, \quad \tilde{N}_{1} > 0, \quad \tilde{H} = \pm 1, \quad \mbox{for Bianchi Type II}.
\end{eqnarray}
We have summarized these shear invariant sets with the corresponding cosmological model and notation in Table ~\ref{table2}, and refer the reader to Page 127 in \cite{ellis} for further details.
\begin{table}[H]
\caption{The various shear invariant sets, with $\alpha = 1,2,3$.}
\begin{center}
\begin{tabular}{|c|c|}
\hline
Notation & Class of Models  \\ \hline
$S_{\alpha}(II) $ & LRS Bianchi II \\ \hline
$S_{\alpha}(VII_{0})$ & LRS Bianchi $VII_{0}$ \\ \hline
$S_{\alpha}(IX)$ & LRS Bianchi $IX$ \\ \hline
\end{tabular}
\end{center}
\label{table2}
\end{table}
Note that in Table ~\ref{table2}, LRS stands for locally rotationally symmetric. Therefore, all the models corresponding to the shear invariant sets are the locally rotationally symmetric Bianchi models. These models are still homogeneous on spacelike orbits, but the dimension of the isotropy subgroup is one greater than the non-locally rotationally symmetric Bianchi models. That is, the LRS Bianchi models belong to the isometry group $G_{4}$ (Pages 22 and 23, \cite{ellis}).

\section{A Fixed-Point Analysis}
In this section we list the equilibrium points of the dynamical system \eqref{syseqs1}-\eqref{syseqs7}. This is an autonomous system, and can be written in the form
\begin{equation}
\dot{x} = v(x), \quad x \in \mathbb{R}^{7}.
\end{equation}
An equilibrium point of the system is a point at which the vector field, $v(x) \in \mathbb{R}^{7}$ vanishes. In our analysis of the stability of these equilibrium points, we first note that an equilibrium point of a differential equation is \emph{hyperbolic} if no eigenvalue of the linear part of the equation at this singular point lies on the imaginary axis (Page 47, \cite{arnolddyn}). We then make use of the Grobman-Hartman theorem (Page 48, \cite{arnolddyn}, Pages 95-96 \cite{ellis}) which says that a $C^{1}$ vector field is topologically equivalent to its linear part in a neighborhood of a hyperbolic equilibrium point. As a consequence of this theorem, if the eigenvalues of the linear part of the system evaluated at the hyperbolic equilibrium point are strictly negative,  the equilibrium point will be a local sink of the system. Similarly, if the eigenvalues of the linear part of the system evaluated at the hyperbolic equilibrium point are strictly positive, the equilibrium point will be a local source of the system. A hyperbolic equilibrium point which is neither a source nor a sink is termed a saddle point. We will then make use of the invariant manifold theorem which will allow us to classify orbits that are either attracted to or repelled by certain hyperbolic equilibrium points as $\tau \to \pm \infty$.

\subsection{Flat Friedmann-LeMa\^{i}tre-Robertson-Walker (FLRW) Equilibrium Points: $F_{\pm}$}
\subsubsection{The Expanding Epoch}
\begin{equation}
\label{FLRWpoints1}
F_{+}: \tilde{\Sigma}_{+} = \tilde{\Sigma}_{-} = 0, \quad \tilde{N}_{1} = \tilde{N}_{2} = \tilde{N}_{3} = 0, \quad \tilde{H} =  1, \quad \tilde{\Omega} = 1
\end{equation}
The eigenvalues are found to be
\begin{equation}
\label{Feigs2}
\lambda_{1} = \lambda_{2} = \lambda_{3} =  \frac{1}{2} \left(1 + 3w - 9\tilde{\xi}_{0}\right), \quad \lambda_{4} = \lambda_{5} = 1 + 3w - 9\tilde{\xi}_{0}, \quad \lambda_{6} = \lambda_{7} = \frac{3}{2} \left(-1 + w - 4 \tilde{\eta}_{0} - 3 \tilde{\xi}_{0}\right).
\end{equation}
$F_{+}$ is a local sink of the system if
\begin{equation}
\label{Fregion1plus}
\tilde{\eta}_{0} \geq 0\wedge \left[\left(0\leq \tilde{\xi}_{0} \leq \frac{4}{9}\wedge-1\leq w<\frac{1}{3} (-1+9 \tilde{\xi}_{0} )\right)\vee \left(\tilde{\xi}_{0} >\frac{4}{9}\wedge-1\leq w\leq 1\right)\right].
\end{equation}

There regions where $F_{+}$ corresponds to a saddle point of the system are:
\begin{equation}
\label{Fregion2plus}
\tilde{\eta}_{0} = 0\wedge \left[\left(\tilde{\xi}_{0} = 0\wedge -\frac{1}{3}<w<1\right)\vee\left(0<\tilde{\xi}_{0} <\frac{4}{9}\wedge\frac{1}{3} (-1+9 \tilde{\xi}_{0} )<w\leq 1\right)\right],
\end{equation}
and
\begin{equation}
\label{Fregion3plus}
\tilde{\eta}_{0} >0\wedge 0\leq \tilde{\xi}_{0} <\frac{4}{9}\wedge \frac{1}{3} (-1+9 \tilde{\xi}_{0} )<w\leq 1.
\end{equation}

We note that there exist no $w, \tilde{\xi}_{0}, \tilde{\eta}_{0}$ corresponding to $-1 \leq w \leq 1$, $\tilde{\xi}_{0}\geq 0$ and $\tilde{\eta}_{0} \geq 0$ such that the eigenvalues presented in Eq. \eqref{Feigs2} are strictly positive. Hence, the equilibrium point, $F_{+}$ corresponding to a flat expanding FLRW solution is not a local source of the system. 


\subsubsection{The Contracting Epoch}
\begin{equation}
\label{FLRWpoints2}
F_{-}: \tilde{\Sigma}_{+} = \tilde{\Sigma}_{-} = 0, \quad \tilde{N}_{1} = \tilde{N}_{2} = \tilde{N}_{3} = 0, \quad \tilde{H} =  -1, \quad \tilde{\Omega} = 1
\end{equation}
The eigenvalues are found to be
\begin{equation}
\label{Feigs1}
\lambda_{1} = \lambda_{2} = \lambda_{3} = \frac{1}{2} \left(-1 -3w - 9 \tilde{\xi}_{0}\right), \quad \lambda_{4} = \lambda_{5} = -\frac{3}{2} \left(-1 + w + 4 \tilde{\eta}_{0} + 3 \tilde{\xi}_{0} \right), \quad \lambda_{6} = \lambda_{7} = -1 - 3w - 9\tilde{\xi}_{0}.
\end{equation}
$F_{-}$ is a local sink of the system in three separate regions of the parameter space. These are given by
\begin{equation}
\label{Fregion1minus}
\tilde{\eta}_{0} = 0 \wedge \left[\left(0<\tilde{\xi}_{0} \leq \frac{2}{3} \wedge 1-3 \tilde{\xi}_{0} <w\leq 1\right) \vee \left(\tilde{\xi}_{0} >\frac{2}{3} \wedge -1\leq w\leq 1\right)\right],
\end{equation}
\begin{equation}
\label{Fregion2minus}
\begin{split}
0<\tilde{\eta}_{0} \leq \frac{1}{3} \wedge \\  \left[\left(0\leq \tilde{\xi}_{0} <\frac{1}{3} (2-4 \tilde{\eta}_{0} )\wedge 1-4 \tilde{\eta}_{0} -3 \tilde{\xi}_{0} <w\leq 1\right)\vee \left(\tilde{\xi}_{0} = \frac{1}{3} (2-4 \tilde{\eta}_{0} ) \wedge-1<w\leq 1\right) \vee \left(\tilde{\xi}_{0} >\frac{1}{3} (2-4 \tilde{\eta}_{0} )\wedge-1\leq w\leq 1\right)\right],
\end{split}
\end{equation}
and
\begin{equation}
\label{Fregion3minus}
\tilde{\eta}_{0} >\frac{1}{3}\wedge \left[\left(0\leq \tilde{\xi}_{0} \leq \frac{2}{9}\wedge \frac{1}{3} (-1-9 \tilde{\xi}_{0} )<w\leq 1\right) \vee \left(\tilde{\xi}_{0} >\frac{2}{9}\wedge-1\leq w\leq 1\right)\right].
\end{equation}

$F_{-}$ is a local source in two separate regions of the parameter space. These are given by
\begin{equation}
\label{Fregion4minus}
0\leq \tilde{\eta}_{0} \leq \frac{1}{3} \wedge 0\leq \tilde{\xi}_{0} <\frac{2}{9} \wedge -1\leq w<\frac{1}{3} (-1-9 \tilde{\xi}_{0} ),
\end{equation}
and
\begin{equation}
\label{Fregion5minus}
\frac{1}{3}<\tilde{\eta}_{0} <\frac{1}{2} \wedge 0\leq \tilde{\xi}_{0} <\frac{1}{3} (2-4 \tilde{\eta}_{0} )\wedge -1\leq w<1-4 \tilde{\eta}_{0} -3 \tilde{\xi}_{0}.
\end{equation}

$F_{-}$ also represents a saddle point if the following regions of the parameter space:
\begin{equation}
\label{Fregion6minus}
\tilde{\eta}_{0} = 0 \wedge \left[\left(\tilde{\xi}_{0} = 0 \wedge -\frac{1}{3}<w<1\right) \vee \left(0<\tilde{\xi}_{0} \leq \frac{2}{9}\wedge \frac{1}{3} (-1-9 \tilde{\xi}_{0} )<w<1-3 \tilde{\xi}_{0} \right) \vee \left(\frac{2}{9}<\tilde{\xi}_{0} <\frac{2}{3}\wedge-1\leq w<1-3 \tilde{\xi}_{0} \right)\right],
\end{equation}

\begin{equation}
\label{Fregion7minus}
0<\tilde{\eta}_{0} <\frac{1}{3} \wedge \left[\left(0\leq \tilde{\xi}_{0} \leq \frac{2}{9} \wedge \frac{1}{3} (-1-9 \tilde{\xi}_{0} )<w<1-4 \tilde{\eta}_{0} -3 \tilde{\xi}_{0} \right) \vee \left(\frac{2}{9}<\tilde{\xi}_{0} <\frac{1}{3} (2-4 \tilde{\eta}_{0} )\wedge-1\leq w<1-4 \tilde{\eta}_{0} -3 \tilde{\xi}_{0} \right)\right],
\end{equation}

\begin{equation}
\label{Fregion8minus}
\frac{1}{3}<\tilde{\eta}_{0} <\frac{1}{2}\wedge \left[\left(0\leq \tilde{\xi}_{0} \leq \frac{1}{3} (2-4 \tilde{\eta}_{0} )\wedge 1-4 \tilde{\eta}_{0} -3 \tilde{\xi}_{0} <w<\frac{1}{3} (-1-9 \tilde{\xi}_{0} )\right)\vee \left(\frac{1}{3} (2-4 \tilde{\eta}_{0} )<\tilde{\xi}_{0} <\frac{2}{9}\wedge-1\leq w<\frac{1}{3} (-1-9 \tilde{\xi}_{0} )\right)\right],
\end{equation}

\begin{equation}
\label{Fregion9minus}
\tilde{\eta}_{0} = \frac{1}{2}\wedge \left[\left(\tilde{\xi}_{0} = 0\wedge -1<w<-\frac{1}{3}\right)\vee\left(0<\tilde{\xi}_{0} <\frac{2}{9}\wedge-1\leq w<\frac{1}{3} (-1-9 \tilde{\xi}_{0} )\right)\right],
\end{equation}
and
\begin{equation}
\label{Fregion10minus}
\tilde{\eta}_{0} >\frac{1}{2}\wedge 0\leq \tilde{\xi}_{0} <\frac{2}{9}\wedge -1\leq w<\frac{1}{3} (-1-9 \tilde{\xi}_{0} ).
\end{equation}

\subsection{Bianchi Type II Equilibrium Points: $B(II)$}

\subsubsection{The Expanding Epoch}
The Bianchi Type II equilibrium point corresponding to the expanding epoch, $\tilde{H} = 1$, shall be denoted by $P_{+}(II)$. This point is given by
\begin{eqnarray}
\label{biipoint1}
\tilde{\Sigma}_{+} &=& \frac{1}{16} \left[17 + 3w + 3 \tilde{\eta}_{0} + 9 w \tilde{\eta}_{0} - \gamma\right], \nonumber \\
\tilde{\Sigma}_{-} &=& 0, \nonumber \\
\tilde{N}_{1} &=& \frac{1}{4} \sqrt{\frac{3}{2}} \left[-3 \left(63 - 38 \tilde{\eta}_{0} - 9 \tilde{\eta}_{0}^2 + 3 \left(w + 3 w \tilde{\eta}_{0}\right)^{2} - 2 w \left(1 -42 \tilde{\eta}_{0} + 9 \tilde{\eta}_{0}^2\right)\right) + \gamma \left(13 + 3w - 9\tilde{\eta}_{0} +9 w \tilde{\eta}_{0}\right) - 288 \tilde{\xi}_{0}\right]^{1/2} \nonumber, \\
\tilde{N}_{2} &=& \tilde{N}_{3} = 0, \tilde{H} = 1, \nonumber \\
 \tilde{\Omega} &=& \frac{1}{32} \left[15 - 3w -54 \tilde{\eta}_{0} - 18 w \tilde{\eta}_{0} - 9 \tilde{\eta}_{0}^2 - 27 w \tilde{\eta}_{0}^2 + \left(1 + 3 \tilde{\eta}_{0}\right) \gamma\right],
\end{eqnarray}
where
\begin{equation}
\label{gammaeq}
\gamma = \left[ \left(17 + 3 \tilde{\eta}_{0} + w \left(3 + 9 \tilde{\eta}_{0}\right)\right)^2 - 64 \left(1 + 3w - 9 \tilde{\xi}_{0}\right)\right]^{1/2}.
\end{equation}
The regions of the parameter space that correspond to this point are
\begin{equation}
\label{b2ineq1}
\tilde{\eta}_{0} = 0 \wedge \left[ \left(\tilde{\xi}_{0} = 0 \wedge -\frac{1}{3} < w < 1 \right) \vee \left(0 < \tilde{\xi}_{0} < \frac{4}{9} \wedge \frac{1}{3} \left(-1 + 9 \tilde{\xi}_{0}\right) < w \leq 1\right)\right],
\end{equation}
and
\begin{equation}
\label{b2ineq2}
\tilde{\eta}_{0} > 0 \wedge 0 \leq \tilde{\xi}_{0} < \frac{4}{9} \wedge \frac{1}{3} \left(-1 + 9 \tilde{\xi}_{0}\right) < w \leq 1.
\end{equation}

Under the first set of inequalities in Eq. \eqref{b2ineq1}, namely, $\tilde{\xi}_{0} = \tilde{\eta}_{0} = 0$, $-1/3 < w < 1$, the point $P_{+}(II)$ takes the form
\begin{equation}
\label{colstew}
\tilde{\Sigma}_{+} = \frac{1}{8} \left(1 + 3w\right), \quad \tilde{\Sigma}_{-} = 0, \quad \tilde{N}_{1} = \frac{3}{4} \left(1 + 2w - 3w^2\right)^{1/2}, \quad \tilde{N}_{2} = \tilde{N}_{3} = 0, \quad \tilde{H} = 1, \quad \tilde{\Omega} = -\frac{3}{16}\left(-5 + w\right),
\end{equation}
where $-\frac{1}{3} < w < 1$.
The eigenvalues corresponding to Eq. \eqref{colstew} are given by
\begin{equation}
\label{colsteweigs}
\lambda_{1} = \frac{3}{2} \left(-1 + w\right), \quad \lambda_{2} = \lambda_{3} = \frac{3}{4} \left(1 + 3 w\right), \quad \lambda_{4} = \lambda_{5} = \left(1 + 3w\right), \quad \lambda_{6} = -\frac{3}{8} \left(2 - 2w + \beta\right), \quad \lambda_{7} = \frac{3}{8} \left(-2 + 2w + \beta\right),
\end{equation}
where
\begin{equation*}
\beta = \left(-6 - 26w + 38 w^2 - 6w^3\right)^{1/2}.
\end{equation*}
By examining these eigenvalues, one sees that the equilibrium point $P_{+}(II)$ can only be a saddle point of the system. The corresponding parameter space region is given by
\begin{equation}
-\frac{1}{3} < w \leq \frac{1}{3} \left(8 - \sqrt{73}\right), \quad \tilde{\xi}_{0} = \tilde{\eta}_{0} = 0.
\end{equation}

To analyze the stability of the equilibrium point as defined in Eq. \eqref{biipoint1} in the rest of the parameter space as given in Eqs. \eqref{b2ineq1} and \eqref{b2ineq2}, we must resort to numerical techniques. The reason is that the characteristic polynomial of the Jacobian matrix in each of the reasons admits roots that cannot be written down in closed form. We conducted a variety of numerical experiments that demonstrated that indeed $P_{+}(II)$ is a saddle point of the system. The results of some of these experiments can be seen in Figs. \ref{fig3} and \ref{fig4}.  

\subsubsection{The Contracting Epoch}
The Bianchi Type II equilibrium point corresponding to the contracting epoch, $\tilde{H} = -1$, which we denote by $P_{-}(II)$ is given by
\begin{eqnarray}
\label{Piipoint2}
\tilde{\Sigma}_{+} &=& \frac{1}{16} \left[-17 - 3w + 3 \tilde{\eta}_{0} + 9 w \tilde{\eta}_{0} + \epsilon\right], \nonumber \\
\tilde{\Sigma}_{-} &=& 0, \nonumber \\
\tilde{N}_{1} &=& \frac{1}{4} \sqrt{\frac{3}{2}} \left[-189 + 6w - 9w^2 - 114 \tilde{\eta}_{0} + 252 w \tilde{\eta}_{0} + 54 w^2 \tilde{\eta}_{0} + 27 \tilde{\eta}_{0}^2 + 54 w \tilde{\eta}_{0}^2 - 81 w^2 \tilde{\eta}_{0}^2 + 288 \tilde{\xi}_{0} + \epsilon\left(13 + 3w + 9 \tilde{\eta}_{0} - 9 w \tilde{\eta}_{0} \right)\right]^{1/2}, \nonumber \\
\tilde{N}_{2} &=& \tilde{N}_{3} = 0, \quad \tilde{H} = -1, \nonumber \\
\tilde{\Omega} &=& \frac{1}{32} \left[15 -3w + 54 \tilde{\eta}_{0} + 18 w \tilde{\eta}_{0} - 9\tilde{\eta}_{0}^2 - 27 w \tilde{\eta}_{0}^2 + \epsilon\left(1-3\tilde{\eta}_{0}\right)\right],
\end{eqnarray}
where
\begin{equation}
\label{eqepsilon}
\epsilon = \left[ \left(-17 + 3 \tilde{\eta}_{0} + w \left(-3 + 9 \tilde{\eta}_{0}\right)\right)^2 - 64 \left(1 + 3w + 9 \tilde{\xi}_{0}\right)\right]^{1/2}.
\end{equation}
There are many regions of the parameter space that correspond to this point. The majority of them are too complex to write out in this paper. As an example, we have listed two of the simpler ones below:
\begin{equation}
\tilde{\eta}_{0} =0\wedge\left(\left(\tilde{\xi}_{0} =0\wedge-\frac{1}{3}<w<1\right)\vee\left(0<\tilde{\xi}_{0} \leq \frac{2}{9}\wedge\frac{1}{3} (-1-9 \tilde{\xi}_{0} )<w<\frac{1}{3} (3-12 \tilde{\xi}_{0} )\right)\vee\left(\frac{2}{9}<\tilde{\xi}_{0} <\frac{1}{2}\wedge-1\leq w<\frac{1}{3} (3-12 \tilde{\xi}_{0} )\right)\right)
\end{equation}

\begin{equation}
0<\tilde{\eta}_{0} <\frac{1}{3} \left(5-2 \sqrt{6}\right)\wedge\left(\left(0\leq \tilde{\xi}_{0} \leq \frac{2}{9}\wedge\frac{1}{3} (-1-9 \tilde{\xi}_{0} )<w<\frac{-3+10 \tilde{\eta}_{0} +9 \tilde{\eta}_{0} ^2+12 \tilde{\xi}_{0} }{-3-6 \tilde{\eta}_{0} +9 \tilde{\eta}_{0} ^2}\right)\vee\left(\frac{2}{9}<\tilde{\xi}_{0} <\frac{1}{6} \left(3-2 \tilde{\eta}_{0} -9 \tilde{\eta}_{0} ^2\right)\wedge-1\leq w<\frac{-3+10 \tilde{\eta}_{0} +9 \tilde{\eta}_{0} ^2+12 \tilde{\xi}_{0} }{-3-6 \tilde{\eta}_{0} +9 \tilde{\eta}_{0} ^2}\right)\right)
\end{equation}
Finding the eigenvalues for this general case is clearly very difficult to do since the characteristic polynomial has no closed-form solutions. However, based on an extensive numerical analysis, we conjecture that this equilibrium point is in fact a saddle, and is hence unstable. We will in fact show later using the method of Chetaev functions that for the case when $0<w<1$ and $\tilde{\xi} \geq 0$, that this point is unstable.

We note that to the best of the authors' knowledge, the Bianchi Type II solutions as presented in Eqs. \eqref{biipoint1} and \eqref{Piipoint2} have not been presented before in the literature, and hence represent new solutions to the Einstein field equations.
Both equilibrium points $P_{\pm}(II)$ belong to the invariant set $S_{\alpha}(II)$ as listed in Table ~\ref{table2}, which corresponds to the class of locally rotationally symmetric Bianchi Type II cosmological models.


\subsection{Kasner Equilibrium Points}
There are two possible Kasner equilibrium points, differing only by the value of $\tilde{H}$:
\begin{equation}
\mathcal{K}_{\pm}: \quad \tilde{\Sigma}_{+}^2 + \tilde{\Sigma}_{-}^2 = 1, \quad \tilde{N}_{1} = \tilde{N}_{2} = \tilde{N}_{3} = 0, \quad \tilde{\Omega} = 0, \quad \quad \tilde{H} = \pm 1,
\end{equation}
where in both cases we have
\begin{equation}
\tilde{\xi}_{0} = \tilde{\eta}_{0} = 0, \quad -1 \leq w < 1.
\end{equation}

Following \cite{whpaper}, we note that the constant values of $\tilde{\Sigma}_{\pm}$ at $\mathcal{K}_{\pm}$ are related to the Kasner exponents of the Kasner solution:
\begin{equation}
\label{kasnerexp}
p_{1} = \frac{1}{3}\left(1 - 2\tilde{\Sigma}_{+}\right), \quad p_{2} = \frac{1}{3}\left(1 + \tilde{\Sigma}_{+} + \sqrt{3}\tilde{\Sigma}_{-}\right), \quad p_{3} = \frac{1}{3}\left(1 + \tilde{\Sigma}_{+} - \sqrt{3}\tilde{\Sigma}_{-}\right).
\end{equation}
\subsubsection{The Expanding Epoch}
The eigenvalues corresponding to $\mathcal{K}_{+}$ are given by
\begin{eqnarray}
\label{kasnereigs1}
\lambda_{1} = \lambda_{2} = 4, \quad \lambda_{3} = 0, \quad \lambda_{4} = 3\left(-1+w\right), \quad \lambda_{5} = 6p_{1}, \quad \lambda_{6} = 6p_{3}, \quad \lambda_{7} = 6p_{2}.
\end{eqnarray}
%
%
The zero eigenvalue in Eq. \eqref{kasnereigs1} indicates that $\mathcal{K}_{+}$ is a one-dimensional family of equilibrium points. Additionally, this zero eigenvalue implies the existence of a one-dimensional center manifold. The Kasner exponents $p_{1}, p_{2}, p_{3}$ obey the relations
\begin{equation}
\label{kasnerrels}
p_{1} + p_{2} + p_{3} = 1, \quad p_{1}^2 + p_{2}^2 + p_{3}^2 = 1,
\end{equation}
which implies that exactly one of $\lambda_{5}, \lambda_{6}$ or $\lambda_{7}$ in Eq. \eqref{kasnereigs1} is negative except when
\begin{equation}
\left(p_{i}\right) = (1,0,0), (0,1,0), (0,0,1) \equiv \left(T_{i}\right), \quad (i = 1,2,3).
\end{equation}
The points $T_{i}$ are the Taub points corresponding to Taub flat spacetime metric (Page 132, \cite{ellis}). We see that in the region $-1 \leq w < 1$, $\lambda_{4} < 0$, and therefore, $\mathcal{K}_{+}$ is a (normally hyperbolic) saddle point. If on the other hand, $w = 1$, then $\lambda_{4} = 0$, which leads to the creation of a two-dimensional center manifold, and the stability behavior in this case cannot be determined by linearization.

\subsubsection{The Contracting Epoch}
The eigenvalues corresponding to $\mathcal{K}_{-}$ are found to be
\begin{eqnarray}
\label{kasnereigs2}
\lambda_{1} &=& \lambda_{2} = -4, \quad \lambda_{3} = 0, \quad \lambda_{4} = 3(-1 + w), \quad \lambda_{5} = -4 + 6p_{1}, \quad \lambda_{6} = -1 -3p_{1} - 3 p_{2} + 3 p_{3}, \nonumber \\ \lambda_{7} &=& -1 - 3p_{1} + 3\sqrt{3}p_{2} - 3 \sqrt{3}p_{3}.
\end{eqnarray}
In general, since the Kasner exponents $p_{i}, (i = 1,2,3)$ must obey the Kasner relations as given in Eq. \eqref{kasnerrels}, $\lambda_{5}, \lambda_{6}$ and $\lambda_{7}$ in Eq. \eqref{kasnereigs2} will  in general have alternating signs. Therefore, in the full state space, $\mathcal{K}_{-}$ is also a saddle point.

\subsection{Jacobs Disc}
We see that two Jacobs disc solutions, corresponding to expanding and contracting epochs are equilibrium points of the system as well:
\begin{equation}
\label{jacobseqpoint}
\mathcal{J}_{\pm}: \tilde{\Sigma}_{+}^2 + \tilde{\Sigma}_{-}^2 < 1, \quad \tilde{N}_{1} = \tilde{N}_{2} = \tilde{N}_{3} = 0, \quad 0 < \tilde{\Omega} < 1, \quad \tilde{H} = \pm 1,
\end{equation}
where $\tilde{\eta}_{0} = \tilde{\xi}_{0} = 0$,  and $w = 1$.
\subsubsection{The Expanding Epoch}
The eigenvalues corresponding to $\mathcal{J}_{+}$ are found to be
\begin{eqnarray}
\label{jacobseigs1}
\lambda_{1} = \lambda_{2} = 4, \quad \lambda_{3} = \lambda_{4} = 0, \quad \lambda_{5} = 6p_{1}, \quad \lambda_{6} = 6 p_{3}, \quad \lambda_{7} = 6 p_{2},
\end{eqnarray}
where $p_{i}, (i = 1,2,3)$ are the Kasner exponents as given in Eq. \eqref{kasnerexp} and satisfy the Kasner relations as given in Eq. \eqref{kasnerrels}. The two zero eigenvalues in Eq. \eqref{jacobseigs1} indicate that $\mathcal{J}_{+}$ is a two-dimensional set of equilibrium points.  As can be shown, the eigenspaces associated with $\lambda_{5}, \lambda_{6}$ and $\lambda_{7}$ in Eq. \eqref{jacobseigs1} are parallel to the $\tilde{N}_{1}, \tilde{N}_{2}$ and $\tilde{N}_{3}$ axes. We can therefore conclude that the subset for which $\lambda_{5,6,7} > 0 $ is a source in the interior of the Kasner circle $\mathcal{K}_{+}$, belonging to the Jacobs disc $\mathcal{J}_{+}$.  

\subsubsection{The Contracting Epoch}
The eigenvalues corresponding to $\mathcal{J}_{-}$ are found to be
\begin{eqnarray}
\label{jacobseigs2}
\lambda_{1} &=& \lambda_{2} = -4, \quad \lambda_{3} = \lambda_{4} = 0, \quad \lambda_{5} = -4 + 6p_{1}, \quad \lambda_{6} = -1 -3p_{1} - 3 p_{2} + 3 p_{3}, \nonumber \\ \lambda_{7} &=& -1 - 3p_{1} + 3\sqrt{3}p_{2} - 3 \sqrt{3}p_{3},
\end{eqnarray}
where $p_{i}, (i = 1,2,3)$ are the Kasner exponents as given in Eq. \eqref{kasnerexp} and satisfy the Kasner relations as given in Eq. \eqref{kasnerrels}. The two zero eigenvalues in Eq. \eqref{jacobseigs2} indicate that $\mathcal{J}_{-}$ is a two-dimensional set of equilibrium points. As in the expanding epoch case, we still have that the eigenspaces associated with $\lambda_{5}, \lambda_{6}$ and $\lambda_{7}$ in Eq. \eqref{jacobseigs2} are parallel to the $\tilde{N}_{1}, \tilde{N}_{2}$ and $\tilde{N}_{3}$ axes. However, we find that $\lambda_{5,6,7} < 0$ in Eq. \eqref{jacobseigs2} if $p_{1} + p_{2} + p_{3} = 1$, that is, the first of the Kasner relations in Eq. \eqref{kasnerrels} is satisfied and
\begin{equation}
-\frac{1}{3} < p_{1} < \frac{2}{3}, \quad p_{1} + p_{2} > \frac{1}{3}, \quad 9 + \sqrt{3} + 3 \left(-3 + \sqrt{3}\right) p_{1} > 18 p_{2}.
\end{equation}
Therefore, we conclude that $\mathcal{J}_{-}$ is a local sink of the system. One can also show that there exists no real values for $p_{1,2,3}$ such that the eigenvalues $\lambda_{5,6,7}$ in Eq. \eqref{jacobseigs2} are greater than zero. Hence, $\mathcal{J}_{-}$ is never a source of the dynamical system.

\subsection{Bianchi Type $VII_{0}$ Equilibrium Points}
\subsubsection{Line of Equilibrium Points originating on $\mathcal{K}_{+}$}
\begin{equation}
\mathcal{L}_{1}^{+}: \tilde{\Sigma}_{+} = -1, \quad \tilde{\Sigma}_{-} = 0, \quad \tilde{N}_{1} = 0, \quad \tilde{N}_{2} = \tilde{N}_{3} = k > 0, \quad \tilde{H} = 1, \quad \tilde{\Omega} = 0, \quad k \in \mathbb{R}, \quad \left(\tilde{\xi}_{0} = \tilde{\eta}_{0} = 0, \quad -1 \leq w < 1\right).
\end{equation}
The eigenvalues corresponding to $\mathcal{L}_{1}^{+}$ are found to be:
\begin{equation}
\label{Lineeigs}
\lambda_{1} = 6, \quad \lambda_{2} = \lambda_{3} = 4, \quad \lambda_{4} = 0, \quad \lambda_{5} = -2 i k, \quad \lambda_{6} = 2 i k, \quad \lambda_{7} = 3 - 3w, 
\end{equation}
where $k  > 0 \in \mathbb{R}$.

In general, we see that $\mathcal{L}_{1}^{+}$ is not hyperbolic because three of its eigenvalues lie entirely on the imaginary axis. In addition, in the case where $-1 \leq w < 1$, there are four eigenvalues that are positive, so that $\mathcal{L}_{1}^{+}$ has a three-dimensional unstable set. However, because of the non-hyperbolic nature of this point, its stability in the full state space cannot be determined by linearization methods. We also note that $\mathcal{L}_{1}^{+}$ is a line of equilibrium points originating from the Taub point $T_{1}$ on $\mathcal{K}_{+}$. We can however, restrict the dynamical system to the shear invariant set $S_{1}(IX)$ as described in Eq. \eqref{shearsets}. Within this shear invariant set, only the eigenvalues $\lambda_{1}, \lambda_{2}, \lambda_{3}, \lambda_{4}$ and $\lambda_{7}$ in Eq. \eqref{Lineeigs} arise. Therefore, we have that within $S_{1}(IX)$, $\mathcal{L}_{1}^{+}$ is a local source.

\subsubsection{Line of Equilibrium Points originating from $F_{\pm}$}
\paragraph{Expanding Epoch}
\begin{eqnarray}
\label{Fexpline}
\mathcal{F}_{1}^{+}(VII_{0})&:& \tilde{\Sigma}_{\pm} = 0, \quad \tilde{N}_{1} = 0, \quad \tilde{N}_{2} = \tilde{N}_{3} = d > 0 \in \mathbb{R}, \quad \tilde{H} = 1, \quad \tilde{\Omega} = 1, \nonumber \\
\mathcal{F}_{2}^{+}(VII_{0})&:& \tilde{\Sigma}_{\pm} = 0, \quad \tilde{N}_{2} = 0, \quad \tilde{N}_{1} = \tilde{N}_{3} = d > 0 \in \mathbb{R}, \quad \tilde{H} = 1, \quad \tilde{\Omega} = 1, \nonumber \\
\mathcal{F}_{3}^{+}(VII_{0})&:& \tilde{\Sigma}_{\pm} = 0, \quad \tilde{N}_{3} = 0, \quad \tilde{N}_{1} = \tilde{N}_{2} = d > 0 \in \mathbb{R}, \quad \tilde{H} = 1, \quad \tilde{\Omega} = 1,
\end{eqnarray}
where the eigenvalues for each point described by Eq. \eqref{Fexpline} are found to be
\begin{eqnarray}
\label{Flineeigs1}
\lambda_{1} &=& \lambda_{2} = \lambda_{3} = \lambda_{4} = 0, \quad \lambda_{5} = -2-6 \tilde{\eta}_{0} , \quad \lambda_{6} = -1-3 \tilde{\eta}_{0} -\sqrt{-4 d^2+(1+3 \tilde{\eta}_{0} )^2}, \nonumber \\ \lambda_{7} &=& -1-3 \tilde{\eta}_{0} +\sqrt{-4 d^2+(1+3 \tilde{\eta}_{0} )^2},
\end{eqnarray}
with
\begin{equation}
\label{paramsFexp}
\tilde{\eta}_{0} \geq 0, \quad 0 \leq \tilde{\xi}_{0} \leq \frac{4}{9}, \quad w = \frac{1}{3}\left(-1 + 9 \tilde{\xi}_{0}\right).
\end{equation}
One can see that from Eq. \eqref{Flineeigs1} the four zero eigenvalues indicate that $\mathcal{F}_{i}^{+} (i = 1,2,3)$ each represent a four-dimensional set of equilibrium points, which imply the existence of a four-dimensional center manifold. It is not possible to determine the stability of these equilibrium points by linearization methods because they are clearly non-hyperbolic. However, an interesting feature to note is that these equilibrium points are only defined in a very specific region of parameter space described by Eq. \eqref{paramsFexp}. Moreover, these lines of equilibrium points originate from the flat equilibrium point $F_{+}$ 
and determine the destabilization of $F_{+}$ at $w = (1/3)\left(-1 + 9 \tilde{\xi}_{0}\right)$.   

\paragraph{Contracting Epoch}
\begin{eqnarray}
\label{Fconline}
\mathcal{F}_{1}^{-}(VII_{0})&:& \tilde{\Sigma}_{\pm} = 0, \quad \tilde{N}_{1} = 0, \quad \tilde{N}_{2} = \tilde{N}_{3} = d > 0 \in \mathbb{R}, \quad \tilde{H} = -1, \quad \tilde{\Omega} = 1, \nonumber \\
\mathcal{F}_{2}^{-}(VII_{0})&:& \tilde{\Sigma}_{\pm} = 0, \quad \tilde{N}_{2} = 0, \quad \tilde{N}_{1} = \tilde{N}_{3} = d > 0 \in \mathbb{R}, \quad \tilde{H} = -1, \quad \tilde{\Omega} = 1, \nonumber \\
\mathcal{F}_{3}^{-}(VII_{0})&:& \tilde{\Sigma}_{\pm} = 0, \quad \tilde{N}_{3} = 0, \quad \tilde{N}_{1} = \tilde{N}_{2} = d > 0 \in \mathbb{R}, \quad \tilde{H} = -1, \quad \tilde{\Omega} = 1,\nonumber \\
\end{eqnarray}
where the eigenvalues for each point described by Eq. \eqref{Fconline} are found to be
\begin{eqnarray}
\label{Flineeigs2}
\lambda_{1} &=& \lambda_{2} = \lambda_{3} = \lambda_{4} = 0, \quad \lambda_{5} = 2-6 \tilde{\eta}_{0} , \quad \lambda_{6} = 1-\sqrt{-4 d^2+(1-3 \tilde{\eta}_{0} )^2}-3 \tilde{\eta}_{0} , \nonumber \\ \lambda_{7} &=& 1+\sqrt{-4 d^2+(1-3 \tilde{\eta}_{0} )^2}-3 \tilde{\eta}_{0},
\end{eqnarray}
with
\begin{equation}
\label{paramsFcon}
\tilde{\eta}_{0} \geq 0, \quad 0 \leq \tilde{\xi}_{0} \leq \frac{2}{9}, \quad w = \frac{1}{3}\left(-1 - 9 \tilde{\xi}_{0}\right).
\end{equation}
One can see that from Eq. \eqref{Flineeigs2} the four zero eigenvalues indicate that $\mathcal{F}_{i}^{-} (i = 1,2,3)$ each represent a four-dimensional set of equilibrium points, which imply the existence of a four-dimensional center manifold. It is also not possible as in the case of the expanding epoch to determine the stability of these equilibrium points by linearization methods because they are clearly non-hyperbolic. Moreover, these lines of equilibrium points originate from the flat equilibrium point $F_{-}$ 
and determine the destabilization  of $F_{-}$ at $w = (1/3)\left(-1 - 9 \tilde{\xi}_{0}\right)$.   

\subsection{Bianchi Type $IX$ Equilibrium Points}
The equilibrium points in the interior of $B(IX)$ are generally described by the following values of the dynamical variables and normalized shear viscosity parameter $\tilde{\eta}_{0}$:
\begin{equation}
\label{eqpointsbix}
F_{c}: \tilde{\Sigma}_{\pm} =0, \quad \tilde{N}_{1} = \tilde{N}_{2} = \tilde{N}_{3} = f > 0 \mathbb{R},  \quad \tilde{\Omega} = 1, \quad \tilde{\eta}_{0} \geq 0,
\end{equation}
where $F_{c}$ denotes a closed FLRW universe, of which the Einstein static universe is a special case.
With the definitions in Eq. \eqref{eqpointsbix}, there are four possibilities involving the values of the other dynamical variables and parameters $w$ and $\tilde{\xi}_{0}$, which we list below in succession. Before continuing, an important point must be made. Upon observing Eq. \eqref{eqpointsbix}, one will notice an \emph{apparent} contradiction between $\tilde{\Omega} = 1$, and that $\tilde{N}_{1} = \tilde{N}_{2} = \tilde{N}_{3} = f > 0,  f\in \mathbb{R}$, since this seems to imply that there is a solution to the Einstein field equations that has constant positive curvature, but with unit matter density. This confusion arises because in standard cosmology theory (which includes the expansion-normalized variables approach to the Bianchi cosmologies based on the theory of orthonormal frames), one relates the curvature of the universe to the matter density in that universe via the generalized Friedmann equation (Page 114, \cite{ellis})
\begin{equation}
\label{Friedmann1}
\Omega = 1 - \Sigma^2 - K,
\end{equation}
which is obtained by normalizing Eq. \eqref{efe3} with the Hubble parameter, $H$ (See \cite{ellis} \cite{whpaper} \cite{elliscosmo} and references therein for further details). In Eq. \eqref{Friedmann1}, $\Omega$ is the expansion-normalized density parameter, $\Sigma^2$ is the expansion-normalized shear scalar parameter, and $K$ is the negated expansion-normalized three-dimensional Ricci scalar. For simplicity, let us assume that the universe we are considering is isotropic so that $\Sigma^2$ vanishes. It is then clear from Eq. \eqref{Friedmann1} that for a positively curved universe, $K < 0$ which implies that $\Omega > 1$. For a negatively curved universe, we have that $K > 0$, which implies that $\Omega < 1$. For a flat universe, $K = 0$, which implies that $\Omega = 1$, it is from this point that the confusion arises. 

In our work, because we have normalized our variables with powers of $D$ and not $H$, we have a slightly different analog of the Friedmann equation as given in Eq. \eqref{modfriedmann}.
Considering the definitions in Eq. \eqref{eqpointsbix}, we have from Eq. \eqref{Vdef} that $\tilde{V} = 0$, and so $\tilde{\Omega} = 1$. However, the three dimensional Ricci scalar as defined in Eq. \eqref{scalars2} for $n_{1} = n_{2} = n_{3} = f > 0, f \in \mathbb{R}$ evaluates to
\begin{equation}
R = \frac{3}{2}f^2,
\end{equation}
which is always positive for $f > 0$. Therefore, the apparent confusion arises due to our choice of normalization variable, and is therefore of no real concern with respect to our analysis of the interior of the $B(IX)$ equilibrium points.  

\subsubsection{Case 1: The Einstein Static Universe}
The line element for the Einstein static universe is given by (Page 55, \cite{ellis})
\begin{equation}
ds^2 = -dt^2 + l^2\left[dr^2 + \sin^2 r \left(d\theta^2 + \sin^2 \theta \phi^2\right)\right],
\end{equation}
where $l > 0$ is a constant. It can be shown that the energy density and pressure corresponding to this line element are given by
\begin{equation}
\mu = \frac{3}{l^2}, \quad p = -\frac{1}{l^2},
\end{equation}
which implies that the equation of state parameter has the value $w = -1/3$. Therefore, it is not necessary to describe the Einstein static universe using a two-fluid description as done in \cite{barrowyama} for example. It is however, the more popular choice to consider two non-interacting fluids that have separate equations of state described by two separate equation of state parameters $w_{1}$ and $w_{2}$. Einstein himself took $w_{1} = 0$, and $w_{2} = -1$, with the latter being equivalent to the cosmological constant (Page 55, \cite{ellis}). However, as discussed by Ellis and Wainwright (Page 55, \cite{ellis}), a spatially homogeneous and isotropic universe with constant positive curvature as we have described in Eq. \eqref{eqpointsbix} with equation of state parameter $w = -1/3$ is indeed the Einstein static universe.

This equilibrium point is described by
\begin{equation}
\label{staticpoint}
\tilde{H} = 0, \quad w = -\frac{1}{3}, \quad f = 2, \quad \tilde{\xi}_{0} \geq 0.
\end{equation}
We find that the eigenvalues are given by
\begin{equation}
\label{eigsix2}
\lambda_{1} = \lambda_{2} = 0, \quad \lambda_{3} = \lambda_{4} = -3\tilde{\eta}_{0} - \sqrt{-8 + 9 \tilde{\eta}_{0}^2}, \quad \lambda_{5} = \lambda_{6} = -3\tilde{\eta}_{0} + \sqrt{-8 + 9 \tilde{\eta}_{0}^2}, \quad \lambda_{7} = \frac{9 \tilde{\xi}_{0}}{2}.
\end{equation}
We note that $\lambda_{3,4,5,6}$ in Eq. \eqref{eigsix2} are strictly negative if and only if $\tilde{\eta}_{0} \geq 2\sqrt{2}/3$. That is, if $\tilde{\eta}_{0} \geq 2 \sqrt{2}/3$, the static universe under consideration admits a four-dimensional stable subset. The stability of the Einstein static universe has been a major topic of study in cosmology ever since Einstein introduced the idea \cite{einstein}. The stability properties were first studied by Lema\^{i}tre \cite{lemaitre1} \cite{lemaitre2} and Eddington \cite{eddington}. More recent studies of the stability of the Einstein static universe were completed by Barrow, Ellis, Maartens and Tsagas \cite{barrowellismaartenstsagas} and Barrow and Yamamoto \cite{barrowyama} as mentioned in the introduction of this paper. 

\subsubsection{Case 2: A Set of Closed FLRW Universes in a Contracting Epoch}
This equilibrium point is described by
\begin{equation}
\label{IXcase1}
-1 < \tilde{H} < 0, \quad 0 \leq \tilde{\xi}_{0} \leq -\frac{2}{9\tilde{H}}, \quad w = \frac{1}{3}\left(-1 + 9 \tilde{H} \tilde{\xi}_{0}\right), \quad f = \sqrt{4 - 4 \tilde{H}^2},
\end{equation}
where there is a unique solution for each value of $\tilde{H}$ for $-1 < \tilde{H} < 0$.
The eigenvalues are found to be
\begin{eqnarray}
\label{eigsix1}
\lambda_{1} &=& \lambda_{2} = 0, \quad \lambda_{3} = \lambda_{4} = -\tilde{H}-3 \tilde{\eta}_{0} -\sqrt{-8+9 \tilde{H}^2+6 \tilde{H} \tilde{\eta}_{0} +9 \tilde{\eta}_{0} ^2}, \quad \lambda_{5} = \lambda_{6} = -\tilde{H}-3 \tilde{\eta}_{0} +\sqrt{-8+9 \tilde{H}^2+6 \tilde{H} \tilde{\eta}_{0} +9 \tilde{\eta}_{0} ^2}, \nonumber \\ \lambda_{7} &=& -\frac{9}{2} \left(-1+\tilde{H}^2\right) \tilde{\xi}_{0}.
\end{eqnarray}
The two zero eigenvalues in Eq. \eqref{eigsix1} indicate that the equilibrium point admits a two-dimensional center manifold. However, because of these two zero eigenvalues, the equilibrium point is non-hyperbolic, and its stability cannot be determined by linearization methods. We also note that there are no values for $\tilde{H}$ and $\tilde{\eta}_{0}$ that satisfy Eqs. \eqref{IXcase1} and \eqref{eqpointsbix} such that $\lambda_{3,4,5,6,7} < 0$ simultaneously. Indeed, $\lambda_{3,4,5,6,7} > 0$ simultaneously if and only if
\begin{equation}
\label{params1case1}
0 < \tilde{\xi}_{0} \leq \frac{2}{9}, \quad -1 < H < 0, \quad \tilde{\eta}_{0} \leq -\frac{H}{3} -\frac{2}{3}\sqrt{2}\sqrt{1-\tilde{H}^2}
\end{equation}
or
\begin{equation}
\label{params2case1}
\tilde{\xi}_{0} > \frac{2}{9}, \quad -\frac{2}{9\tilde{H}} \leq \tilde{H} < 0, \quad \tilde{\eta}_{0} \leq -\frac{\tilde{H}}{3} - \frac{2}{3}\sqrt{2}{\sqrt{1-\tilde{H}^2}}.
\end{equation}
Therefore, there is strong evidence to suggest that if one could find some subset of the domain $\mathbb{R}^{7}$ such that $\lambda_{3,4,5,6,7} > 0$ while satisfying the parameter conditions in Eqs. \eqref{params1case1} and \eqref{params2case1}, then this equilibrium point would represent a local source at least within this subset.


\subsubsection{Case 3: A Set of Closed FLRW Universes in an Expanding Epoch}
This equilibrium point is described by
\begin{equation}
\label{IXcase3}
0 < \tilde{H} < 1, \quad 0 \leq \tilde{\xi}_{0} < \frac{4}{9\tilde{H}}, \quad w = \frac{1}{3}\left(-1 + 9 \tilde{H} \tilde{\xi}_{0}\right), \quad f = \sqrt{4 - 4\tilde{H}^2},
\end{equation}
where there is a unique solution for each value of $\tilde{H}$ for $0 < \tilde{H} < 1$.
The eigenvalues are found to be
\begin{eqnarray}
\label{eigsix3}
\lambda_{1} &=& \lambda_{2} = 0, \quad \lambda_{3} = \lambda_{4} = -\tilde{H}-3 \tilde{\eta}_{0} -\sqrt{-8+9 \tilde{H}^2+6 \tilde{H} \tilde{\eta}_{0} +9 \tilde{\eta}_{0} ^2}, \quad \lambda_{5} = \lambda_{6} = -\tilde{H}-3 \tilde{\eta}_{0} +\sqrt{-8+9 \tilde{H}^2+6 \tilde{H} \tilde{\eta}_{0} +9 \tilde{\eta}_{0} ^2}, \nonumber \\ \lambda_{7} &=& -\frac{9}{2} \left(-1+\tilde{H}^2\right) \tilde{\xi}_{0}.
\end{eqnarray}
The two zero eigenvalues in Eq. \eqref{eigsix3} indicate that the equilibrium point has associated with it a two-dimensional center manifold. However, because of these two zero eigenvalues, the equilibrium point is non-hyperbolic, and its stability cannot be determined by linearization methods. We also note that there are no values for $\tilde{H}$ and $\tilde{\eta}_{0}$ that satisfy Eqs. \eqref{IXcase3} and \eqref{eqpointsbix} such that $\lambda_{3,4,5,6,7} < 0$ or $\lambda_{3,4,5,6,7} > 0$ simultaneously. 

\subsubsection{Case 4: A Set of Closed FLRW Universes in an Expanding Epoch}
This equilibrium point is described by
\begin{equation}
\label{IXcase4}
0 < \tilde{H} < 1, \quad \tilde{\xi}_{0} = \frac{4}{9\tilde{H}}, \quad w = 1, \quad f = \sqrt{4 - 4\tilde{H}^2},
\end{equation}
where there is a unique solution for each value of $\tilde{H}$ for $0 < \tilde{H} < 1$. This equilibrium point only arises for $w = 1$, which corresponds to a stiff fluid.
The eigenvalues are found to be
\begin{eqnarray}
\label{eigsix4}
\lambda_{1} &=& \lambda_{2} = 0, \quad \lambda_{3} = \frac{2}{\tilde{H}} - 2 \tilde{H}, \quad \lambda_{4} = \lambda_{5} = -H-3 \tilde{\eta}_{0} -\sqrt{-8+9 \tilde{H}^2+6 \tilde{H} \tilde{\eta}_{0} +9 \tilde{\eta}_{0} ^2}, \nonumber \\ \lambda_{6} &=& \lambda_{7} = -\tilde{H}-3 \tilde{\eta}_{0} +\sqrt{-8+9 \tilde{H}^2+6 \tilde{H} \tilde{\eta}_{0} +9 \tilde{\eta}_{0} ^2}.
\end{eqnarray}
The two zero eigenvalues in Eq. \eqref{eigsix4} indicate that the equilibrium point has associated with it a two-dimensional center manifold. However, because of these two zero eigenvalues, the equilibrium point is non-hyperbolic, and its stability cannot be determined by linearization methods. We also note that there are no values for $\tilde{H}$ and $\tilde{\eta}_{0}$ that satisfy Eqs. \eqref{IXcase4} and \eqref{eqpointsbix} such that $\lambda_{3,4,5,6,7} < 0$ or $\lambda_{3,4,5,6,7} > 0$ simultaneously. 

\subsection{Global Behavior}
Complementing the preceding fix-point analysis, we wish to obtain some information about the asymptotic behavior of the dynamical system as $\tau \to \pm \infty$. To accomplish this, we make use of both the LaSalle Invariance Principle and Monotonicity Principle.  According to Theorem 4.11 in \cite{ellis}, the LaSalle Invariance Principle for $\omega$-limit sets is stated as follows. Consider a dynamical system $\mathbf{x}' = \mathbf{f(x)}$ on $\mathbb{R}^{n}$, with flow $\phi_{t}$. Let $S$ be a closed, bounded and positively invariant set of $\phi_{t}$ and let $Z$ be a $C^{1}$ monotone function. Then $\forall$ $\mathbf{x}_{0} \in S$, we have that $\omega(\mathbf{x}_{0}) \subseteq \left\{\mathbf{x} \in S | Z' = 0\right\}$, where $Z' = \nabla Z \cdot \mathbf{f}$. The extended LaSalle Invariance Principle for $\alpha$-limit sets can be found in Proposition B.3. in \cite{hewwain}. To use this principle, one simply considers $S$ to be a closed, bounded, and negatively invariant set. Then $\forall$ $\mathbf{x}_{0} \in S$, we have that $\alpha(\mathbf{x}_{0}) \subseteq \left\{\mathbf{x} \in S | Z' = 0\right\}$, where $Z' = \nabla Z \cdot \mathbf{f}$. 

The Monotonicity Principle (Proposition A1, \cite{leblanc3}) says if $\phi_{t}$ is a flow on $\mathbb{R}^{n}$ with $S$ an invariant set, and if $Z: S \to \mathbb{R}$ is a $C^{1}$ function whose range is the interval $(a,b)$, where $a \in \mathbb{R} \cup \{-\infty\}$, $b \in \mathbb{R} \cup \{+\infty\}$ and $a < b$, then if $Z$ is monotone decreasing on orbits in $S$, for all $\mathbf{x} \in S$ we have that $\omega(\mathbf{x}) \subseteq \left\{\mathbf{s} \in \bar{S} \backslash S: \lim_{\mathbf{y} \to \mathbf{s}} Z (\mathbf{y}) \neq  b\right\}$,  $\alpha(\mathbf{x}) \subseteq \left\{\mathbf{s} \in \bar{S} \backslash S: \lim_{\mathbf{y} \to \mathbf{s}} Z (\mathbf{y}) \neq  a\right\}$.

Following pages 24 and 25 of \cite{arnolddyn}, we note that a differentiable function $Z$ is called a \emph{Chetaev function} for a singular point $\mathbf{x_{0}}$ of a vector field $\mathbf{f(x)}$ if $Z$ is defined on a domain $W$ whose boundary contains $\mathbf{x}_{0}$, the part of the boundary of $W$ is strictly contained in a sufficiently small ball with its center $\mathbf{x}_{0}$ removed is a piecewise-smooth, $C^{1}$ hypersurface along which $\mathbf{f(x)}$ points into the interior of the domain, that is,
\begin{equation}
Z(\mathbf{x}) \to 0, \mbox{ as } \mathbf{x} \to \mathbf{x}_{0}, \quad \mathbf{x} \in W; \quad Z > 0, \quad \nabla Z \cdot \mathbf{f(x)} > 0 \in W.
\end{equation}
A singular point of a $C^{1}$ vector field for which a Chetaev function exists is unstable.

Let us first consider the function
\begin{equation}
Z_{1} = \tilde{\Omega}.
\end{equation}
Upon using Eqs. \eqref{syseqs7}, \eqref{qdef}, \eqref{modfriedmann} and \eqref{Vdef} , we see that
\begin{equation}
\label{z1ev}
Z_{1}' = -\tilde{\eta}_{0} \left(-12 + \delta + 12 \tilde{\Omega}\right) - \frac{1}{3}\tilde{H}\left[27 \tilde{H} \tilde{\xi}_{0} \left(-1 + \tilde{\Omega}\right) + \tilde{\Omega} \left(-9 + \delta + 9w + 9\tilde{\Omega} - 9w\tilde{\Omega}\right)\right],
\end{equation}
where
\begin{equation}
\delta =  \tilde{N}_{1}^2 + \tilde{N}_{2}^2 - 2\tilde{N}_{2} \tilde{N}_{3} + \tilde{N}_{3}^2 + 3 \left(\tilde{N}_{1} \tilde{N}_{2} \tilde{N}_{3}\right)^{2/3} - 2 \tilde{N}_{1} \left(\tilde{N}_{2} + \tilde{N}_{3}\right).
\end{equation}
In the Bianchi I invariant set $B(I)$ with $\tilde{H} = 1$, Eq. \eqref{z1ev} becomes
\begin{equation}
Z_{1}' = -3 \left(-1 + \tilde{\Omega}\right) \left[4 \tilde{\eta}_{0} + \left(3 \tilde{H} \tilde{\xi}_{0} + \tilde{\Omega} - w \tilde{\Omega}\right)\right].
\end{equation}
This is monotone decreasing in two cases. First, if $\tilde{\eta}_{0} = \tilde{\xi}_{0} = 0$, then $Z_{1}$ is monotone decreasing if
\begin{equation}
\left(w = 1,  0 < \tilde{\Omega} < 1\right) \vee \left(\tilde{\Omega} = 0, -1 \leq w \leq 1 \right).
\end{equation}
Second, in the general viscous case where $\tilde{\xi}_{0} \geq 0, \tilde{\eta}_{0} \geq 0$, $Z_{1}$ is monotone decreasing if
\begin{equation}
-1 \leq w \leq 1,  \quad \tilde{\Omega} \geq 1.
\end{equation}
On the other hand, considering the $B(I)$ set with $\tilde{H} = -1$, Eq. \eqref{z1ev} becomes
\begin{equation}
Z_{1}' = -3 \left(-1 + \tilde{\Omega}\right) \left[4 \tilde{\eta}_{0} + 3 \tilde{\xi}_{0} + \left(-1 + w \right) \tilde{\Omega}\right].
\end{equation}
This is monotone decreasing if
\begin{equation}
\tilde{\xi}_{0} = 0, \quad \tilde{\eta}_{0} = 0, \quad 0 \leq \tilde{\Omega} \leq 1, \quad -1 \leq w \leq 1.
\end{equation}
By the LaSalle invariance principle and the preceding fixed-point analysis, we conclude that for any orbit $\Gamma$
\begin{equation}
\alpha(\Gamma) \in F_{-} \in B(I), \quad \omega(\Gamma) \in F_{+} \in B(I).
\end{equation}
We can also conclude that in the expanding epoch, where $\tilde{H} = 1$, 
\begin{equation}
\alpha(\Gamma) \in \mathcal{J}_{+} \in B(I), \quad \tilde{\eta}_{0} = \tilde{\xi} = 0, w = 1.
\end{equation}
For the contracting epoch, where $\tilde{H} = -1$, we have that
\begin{equation}
\omega(\Gamma) \in \mathcal{J}_{-} \in B(I), \quad \tilde{\eta}_{0} = \tilde{\xi} = 0, w = 1.
\end{equation}

Let us now consider the function
\begin{equation}
Z_{2} = \left(\tilde{N}_{1} \tilde{N}_{2} \tilde{N}_{3}\right)^{2}
\end{equation}
as suggested on Page 149 of \cite{ellis}. Upon using Eqs. \eqref{syseqs4}, \eqref{syseqs5}, \eqref{syseqs6} and \eqref{qdef}, we see that
\begin{equation}
\label{Z2mod}
Z_{2}' = 3 \tilde{H} Z_{2} \left(-9 \tilde{H} \tilde{\xi}_{0} + 4 \hat{\Sigma}^2 + \tilde{\Omega} + 3 w \tilde{\Omega}\right),
\end{equation}
Therefore, $Z_{2}$ is strictly monotone decreasing in the invariant set
\begin{equation}
S_{1} = \left\{\mathbf{x}: -1 < \tilde{H} < 0 \wedge \tilde{N}_{1,2,3} > 0 \wedge \hat{\Sigma}^2 > 0 \wedge \tilde{\Omega} > 0 \right\},
\end{equation}
where $-1/3 \leq w \leq 1$.
The boundary of $S_{1}$ is the invariant set
\begin{equation}
\bar{S}_{1} \backslash S_{1} = \left\{\mathbf{x}: \tilde{H} = -1   \right\} \cup \left\{\mathbf{x}: \tilde{H} = 0\right\} \cup \left\{\mathbf{x}:\tilde{N}_{1,2,3} = 0\right\} \cup \left\{\mathbf{x}: \tilde{\Sigma}_{+}^2 + \tilde{\Sigma}_{-}^2 = 0\right\} \cup \left\{\mathbf{x}: \tilde{\Omega} = 0\right\}.
\end{equation}
Therefore, by the monotonicity principle,
\begin{equation}
\omega \left(\mathbf{x}\right) \subseteq \left\{\mathbf{x}: \tilde{N}_{1,2,3} = 0\right\}
\end{equation}
What this shows is that the future asymptotic state of the $B(IX)$ invariant set belongs to the set $\left\{\mathbf{x}: \tilde{N}_{1,2,3} = 0\right\}$, which according to our fixed-point analysis can either correspond to the Jacobs disc $\mathcal{J}_{-}$ or the flat FLRW universe $F_{-}$.

Considering the function 
\begin{equation}
Z_{3} = \tilde{\Sigma}_{-}^2,
\end{equation}
which was suggested as a monotone function on orbits of $B(II)$ by Wainwright (Page 150, \cite{ellis}).
Upon using Eqs. \eqref{qdef}, \eqref{syseqs3},  \eqref{modfriedmann} and \eqref{Vdef}, and restricting to the $B(II)$ invariant set (with $\tilde{H} = 1$), we see that
\begin{equation}
Z_{3}' = Z_{3} \left[-12 \tilde{\eta}_{0} - 9 \tilde{\xi}_{0} - \frac{1}{3}\tilde{N}_{1}^2 - 3\tilde{\Omega}\left(1-w\right)\right].
\end{equation}
Therefore, $Z_{3}$ is strictly monotone decreasing in the invariant set
\begin{equation}
S_{2} = \left\{\mathbf{x}: \tilde{\Sigma}_{-} > 0  \wedge \tilde{N}_{1} > 0  \wedge  \tilde{\Omega} > 0\right\}.
\end{equation}
The boundary of $S_2$ is the invariant set
\begin{equation}
\bar{S_{2}} \backslash S_{2} = \left\{\mathbf{x}: \tilde{\Sigma}_{-} = 0\right\} \cup \left\{\mathbf{x}: \tilde{N}_{1} = 0\right\} \cup \left\{\mathbf{x}: \tilde{\Omega} = 0 \right\}.
\end{equation}
Therefore, by the monotonicity principle,
\begin{eqnarray}
\omega \left(\mathbf{x}\right) &\subseteq& \left\{\mathbf{x}: \tilde{\Sigma}_{-} = 0\right\}, \\
\alpha  \left(\mathbf{x}\right) &\subseteq& \left\{\mathbf{x}: \tilde{N}_{1} = 0, \tilde{\Omega} = 0 \right\}.
\end{eqnarray} 
This result shows that although in the full state space the point $P_{+}(II)$ according to our fixed point analysis represented a saddle point, restricting to the Bianchi II shear invariant set, the point $P_{+}(II)$ is a local sink by the monotonicity principle. 

%

Finally, let us consider the function
\begin{equation}
Z_{4} = \frac{(1+\tilde{H})^2}{(1-\tilde{H})^2}.
\end{equation}
We will define a domain $W$ as
\begin{eqnarray}
W &=& \nonumber \\
\tilde{\Sigma}_{+} &>& \frac{1}{16} \left[-17 - 3w + 3 \tilde{\eta}_{0} + 9 w \tilde{\eta}_{0} + \epsilon\right] \cup \nonumber \\
\tilde{\Sigma}_{-} &>& 0 \cup \nonumber \\
\tilde{N}_{1} &>& \frac{1}{4} \sqrt{\frac{3}{2}} \left[-189 + 6w - 9w^2 - 114 \tilde{\eta}_{0} + 252 w \tilde{\eta}_{0} + 54 w^2 \tilde{\eta}_{0} + 27 \tilde{\eta}_{0}^2 + 54 w \tilde{\eta}_{0}^2 - 81 w^2 \tilde{\eta}_{0}^2 + 288 \tilde{\xi}_{0} + \epsilon\left(13 + 3w + 9 \tilde{\eta}_{0} - 9 w \tilde{\eta}_{0} \right)\right]^{1/2} \cup \nonumber \\
\tilde{N}_{2} &>& 0 \cup \tilde{N}_{3} > 0 \cup \tilde{H} < -1 \cup \nonumber \\
\tilde{\Omega} &>&  \frac{1}{32} \left[15 -3w + 54 \tilde{\eta}_{0} + 18 w \tilde{\eta}_{0} - 9\tilde{\eta}_{0}^2 - 27 w \tilde{\eta}_{0}^2 + \epsilon\left(1-3\tilde{\eta}_{0}\right)\right],
\end{eqnarray}
where $\epsilon$ is given in Eq. \eqref{eqepsilon}.
The boundary of this domain, $\partial W$ then contains the equilibrium point $P_{-}(II)$ as described in Eq. \eqref{Piipoint2}. One can then show that
\begin{equation}
Z_{4}' = -\frac{\left(1 + \tilde{H}\right)  \left(9 \tilde{H} \tilde{\xi}_{0} - 4 \hat{\Sigma}^2 - \tilde{\Omega} - 3w \tilde{\Omega}\right)}{-1+\tilde{H}}.
\end{equation}
Therefore, we have that
\begin{equation}
Z_{4}(\mathbf{x}) \to 0, \mbox{ as } \mathbf{x} \to P_{-}(II), \quad \mathbf{x} \in W; \quad Z_{4} > 0, \quad \nabla Z_{4} \cdot \mathbf{f(x)} > 0 \in W, \quad \tilde{\xi}_{0} > 0, \quad 0 < w < 1. 
\end{equation}
This implies that for the case when $\tilde{\xi}_{0} > 0, \quad 0 < w < 1$, the point $P_{-}(II)$ is unstable. 

\subsection{Bifurcations}
The dynamical system under study admits some bifurcations. That is, some of the equilibrium points found above change their stability behavior for different values of the equation of state parameter $w$, and the bulk and shear viscosity parameters $\tilde{\xi}_{0}$ and $\tilde{\eta}_{0}$.  A local bifurcation occurs when the Jacobian matrix of the corresponding equilibrium point has at least one eigenvalue with zero real part. If this eigenvalue lies entirely on the real axis, the bifurcation is known as a steady-state bifurcation.

With respect to local bifurcations, we consider only hyperbolic equilibrium points.
Looking first at the equilibrium point $F_{+}$, local bifurcations can occur if
\begin{equation}
0 \leq \tilde{\xi}_{0} \leq \frac{4}{9}, \quad w = \frac{1}{3}\left(-1+9\tilde{\xi}_{0}\right)
\end{equation}
or
\begin{equation}
\tilde{\xi}_{0} = 0, \quad \tilde{\eta}_{0} = 0, \quad w = 1.
\end{equation}

At the equilibrium point $F_{-}$, local bifurcations can occur if
\begin{equation}
\tilde{\eta}_{0} \geq 0, \quad 0 \leq \tilde{\xi}_{0} \leq \frac{2}{9}, \quad w = \frac{1}{3}\left(-1 - 9\tilde{\xi}_{0}\right),
\end{equation}
or
\begin{equation}
\tilde{\xi}_{0} = 0, \quad \tilde{\eta}_{0} = 0, \quad w = 1.
\end{equation}

The only other purely hyperbolic points are $P_{\pm}(II)$, which as can be seen from the corresponding eigenvalues, never admit local bifurcations.

\subsection{Heteroclinic Orbits}
From the preceding fixed-points analysis, one can obtain information about heteroclinic orbits produced by the dynamical system. Heteroclinic orbits are simply orbits that connect \emph{distinct} equilibrium points. 
A very interesting heteroclinic orbit is generated via the equilibrium points $\mathcal{J}_{\pm}$. In the interior of $\mathcal{K}_{\pm}$, we have
\begin{displaymath}
\xymatrix{\mathcal{J}_{+} \ar@{<->}[r] & \mathcal{J}_{-}}
\end{displaymath}
where $\mathcal{J}_{+}$ was found to be a local source within $\mathcal{K}_{+}$ and $\mathcal{J}_{-}$ was found to be a local sink within $\mathcal{K}_{-}$.

Another interesting heteroclinic orbit is one that connects the equilibrium points $\mathcal{K}_{\pm}$. That is,
\begin{displaymath}
\xymatrix{\mathcal{K}_{+} \ar@{<->}[r] & \mathcal{K}_{-}}
\end{displaymath}

We can also have
\begin{displaymath}
\xymatrix{
F_{-} \ar[d] \ar[r] & \mathcal{K}_{-} \ar@{<->}[ld] \\
\mathcal{K}_{+}
}
\end{displaymath}
where $F_{-}$ is a local source, and $\mathcal{K}_{\pm}$ are saddle points.

\subsection{Mixmaster Attractor}
We now briefly describe the famous Mixmaster attractor that generically appears in the study of the dynamics of $B(IX)$ models. One typically observes very complex dynamical behavior, albeit chaotic behavior as such models are evolved in the past towards $\mathcal{K}_{\pm}$. As our fixed point analysis demonstrated, one can only hope to evolve towards $\mathcal{K}_{\pm}$ if $\tilde{\eta}_{0} = \tilde{\xi}_{0} = 0$. There are numerous studies in the literature of Mixmaster dynamics and chaotic behavior in perfect-fluid $B(IX)$ models, many of which we have mentioned in the introduction of this paper. The interested reader should refer to the papers listed there for further elaboration on the points we make in this subsection. 

We begin by noting that in the interior of $B(IX)$, there exists no equilibrium point that is a well-defined local source. Following Section 6.4 and the references therein in \cite{ellis}, we attempt to construct a \emph{compact} invariant set in $\overline{B(IX)} = B(IX) \cup \partial B(IX)$ that is conjectured to be a past attractor. We know from our analysis of the point $\mathcal{K}_{+}$, that there are six families of Taub orbits. Let us consider Eqs. \eqref{modfriedmann} and \eqref{Vdef} in the vacuum boundary, $\tilde{\Omega} = 0$. Each family lies on a half-ellipsoid. The  closures of these six half-ellipsoids are defined as
\begin{eqnarray}
E_{1}^{+}&:& \tilde{\Sigma}_{+}^2+ \tilde{\Sigma}_{-}^2 + \frac{1}{12} \tilde{N}_{1}^2 = 1, \quad \tilde{N}_{1} > 0, \quad \tilde{N}_{2} = \tilde{N}_{3} = 0, \\
E_{1}^{-} &:&  \tilde{\Sigma}_{+}^2+ \tilde{\Sigma}_{-}^2 + \frac{1}{12} \tilde{N}_{1}^2 = 1, \quad \tilde{N}_{1} < 0, \quad \tilde{N}_{2} = \tilde{N}_{3} = 0, \\
E_{2}^{+} &:& \tilde{\Sigma}_{+}^2+ \tilde{\Sigma}_{-}^2 + \frac{1}{12} \tilde{N}_{2}^2 = 1, \quad \tilde{N}_{2} > 0, \quad \tilde{N}_{1} = \tilde{N}_{3} = 0, \\
E_{2}^{-} &:& \tilde{\Sigma}_{+}^2+ \tilde{\Sigma}_{-}^2 + \frac{1}{12} \tilde{N}_{2}^2 = 1, \quad \tilde{N}_{1} < 0, \quad \tilde{N}_{1} = \tilde{N}_{3} = 0, \\
E_{3}^{+} &:& \tilde{\Sigma}_{+}^2+ \tilde{\Sigma}_{-}^2 + \frac{1}{12} \tilde{N}_{3}^2 = 1, \quad \tilde{N}_{3} > 0, \quad \tilde{N}_{2} = \tilde{N}_{1} = 0, \\
E_{3}^{-} &:& \tilde{\Sigma}_{+}^2+ \tilde{\Sigma}_{-}^2 + \frac{1}{12} \tilde{N}_{3}^2 = 1, \quad \tilde{N}_{3} < 0, \quad \tilde{N}_{2} = \tilde{N}_{1} = 0.
\end{eqnarray}
The Taub orbits $T_{i},  (i = 1,2,3)$ and equilibrium points on $\mathcal{K}_{+}$ imply the existence of infinite heteroclinic sequences that map $\mathcal{K}_{+}$ onto itself. The chaotic dynamical behavior can be seen from the basic notions that first, $\overline{B(IX)}$ is conjectured to be an attractor. Second, since orbits have to be confined within the region of the attractor and $\mathcal{K}_{+}$ is a saddle, orbits will indefinitely leave $\mathcal{K}_{+}$ and then approach $\mathcal{K}_{+}$ via the Taub points. This suggests that in the case where $-1 \leq w < 1$, and $\tilde{\xi}_{0} = \tilde{\eta}_{0} = 0$, as $\tau \to -\infty$ the union of $\mathcal{K}_{+}$ and the family of Taub orbits is the past attractor of the dynamical system. Specifically, we have that
\begin{equation}
M^{+}: E_{1}^{+} \cup E_{2}^{+} \cup E_{3}^{+},
\end{equation}
where $M^{+}$ denotes the \emph{Mixmaster attractor}. Let us additionally define the scalar quantity
\begin{equation}
\label{Deltadef}
\Delta = \left(\tilde{N}_{1} \tilde{N}_{2}\right)^{2} + \left(\tilde{N}_{2} \tilde{N}_{3}\right)^{2} + \left(\tilde{N}_{3} \tilde{N}_{1}\right)^{2}.  
\end{equation}
Showing that $M^{+}$ is indeed an attractor requires one to show that both $\Delta$ and $\tilde{\Omega}$ vanish as $\tau \to -\infty$. In the next section, we perform some numerical experiments to test this hypothesis.

\section{Numerical Solutions}
In this section, we perform numerical experiments to complement the analysis of the dynamical system in the previous sections. For all of the numerical experiments, initial conditions denoted by asterisks in the figures were chosen such that Eqs. \eqref{constr1} and \eqref{modfriedmann} were satisfied. The goal of this section is to complement the preceding stability analysis of the equilibrium points with extensive numerical experiments in order to confirm that the local results are in fact global in nature. The details of the parameters used are described in the captions of the respective figures and are based on the fixed-point analysis that characterized the stability of the equilibrium points in the different regions of the parameter space $\left\{\tilde{\eta}_{0}, \tilde{\xi}_{0}, w: \tilde{\eta}_{0} \geq 0, \tilde{\xi}_{0} \geq 0, -1 \leq w \leq 1\right\}$.

\begin{figure}[H]
\begin{center}
\caption{This figure shows the dynamical system behavior for $\tilde{\xi}_{0} = 0$, $\tilde{\eta}_{0} = 1/2 $, and $w = -1/3$. The plus sign denotes the equilibrium point $F_{+}$. The model also isotropizes as can be seen from the last figure, where $\tilde{\Sigma}_{\pm} \to 0$ as $\tau \to \infty$. Numerical solutions were computed for $0 \leq \tau \leq 1000$. For clarity, we have displayed solutions for shorter timescales.} 
\label{fig1}
\includegraphics*[scale = 0.35]{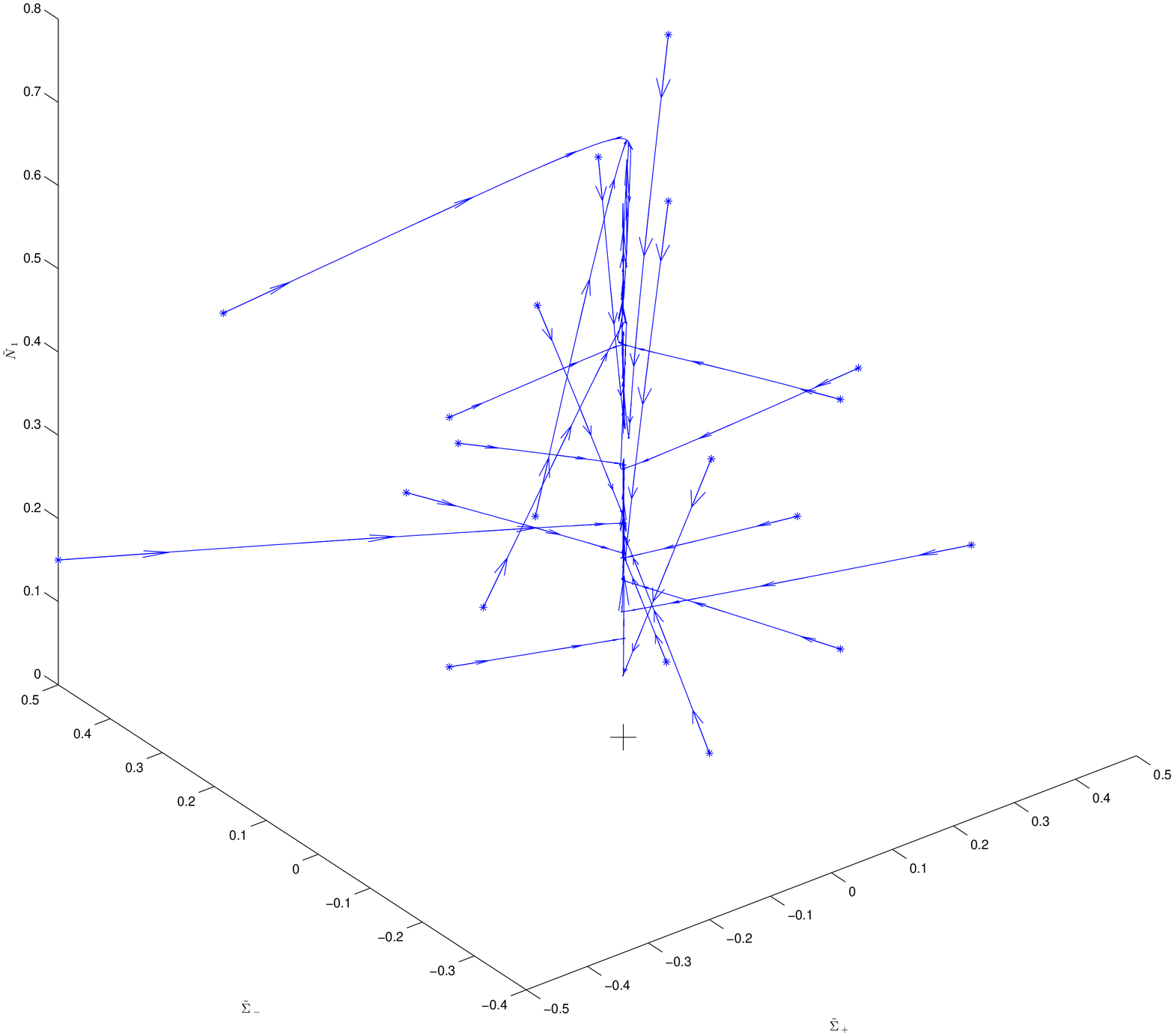}
\includegraphics*[scale = 0.35]{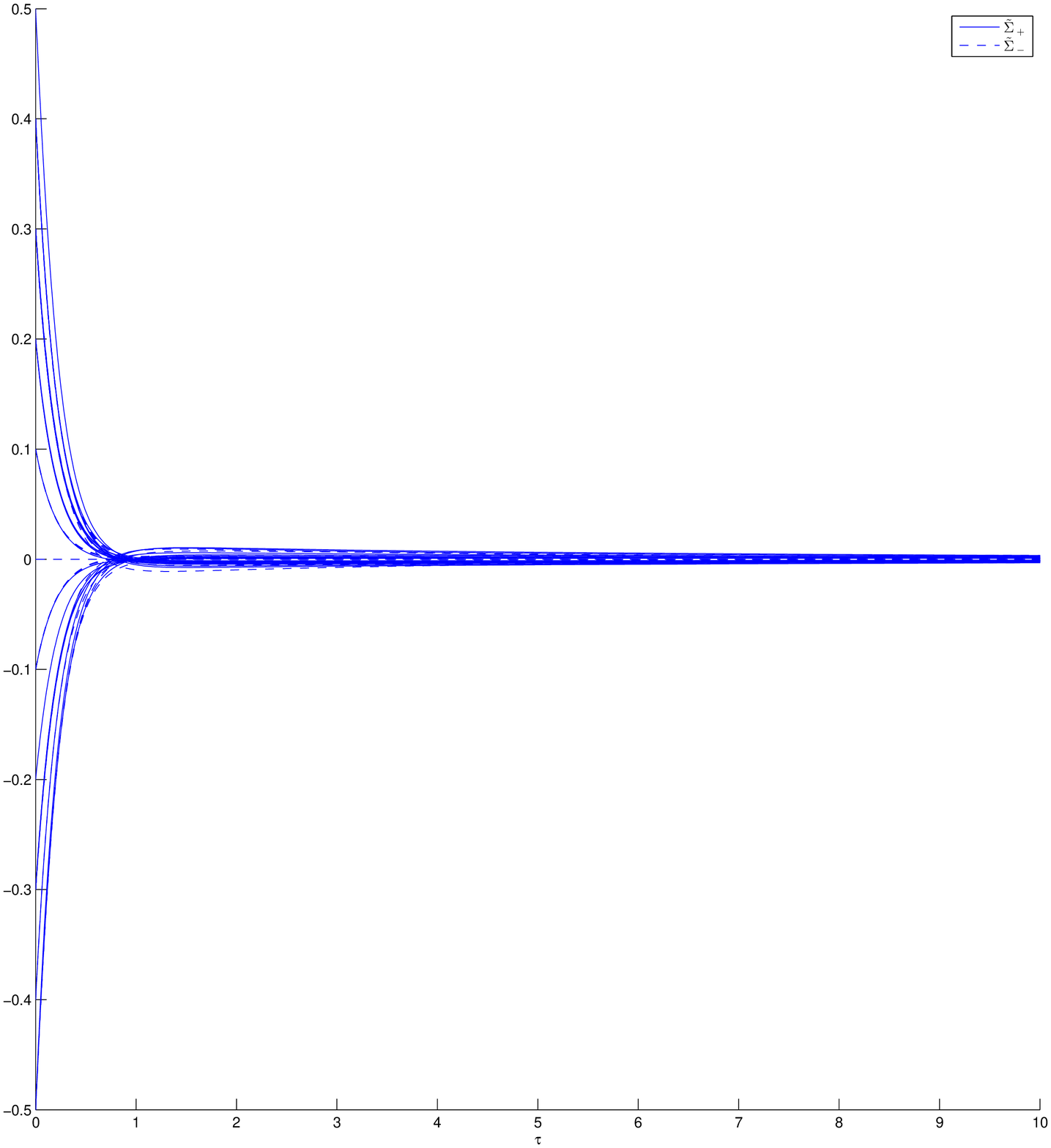}
\end{center}
\end{figure}

\begin{figure}[H]
\begin{center}
\caption{This figure shows the dynamical system behavior for $\tilde{\xi}_{0} = 0$, $\tilde{\eta}_{0} = 1/3$, and $w = 1$. The plus sign denotes the equilibrium point $F_{-}$. The model also isotropizes as can be seen from the last figure, where $\tilde{\Sigma}_{\pm} \to 0$ as $\tau \to \infty$. Numerical solutions were computed for $0 \leq \tau \leq 1000$. For clarity, we have displayed solutions for shorter timescales.} \label{fig2}
\includegraphics*[scale = 0.45]{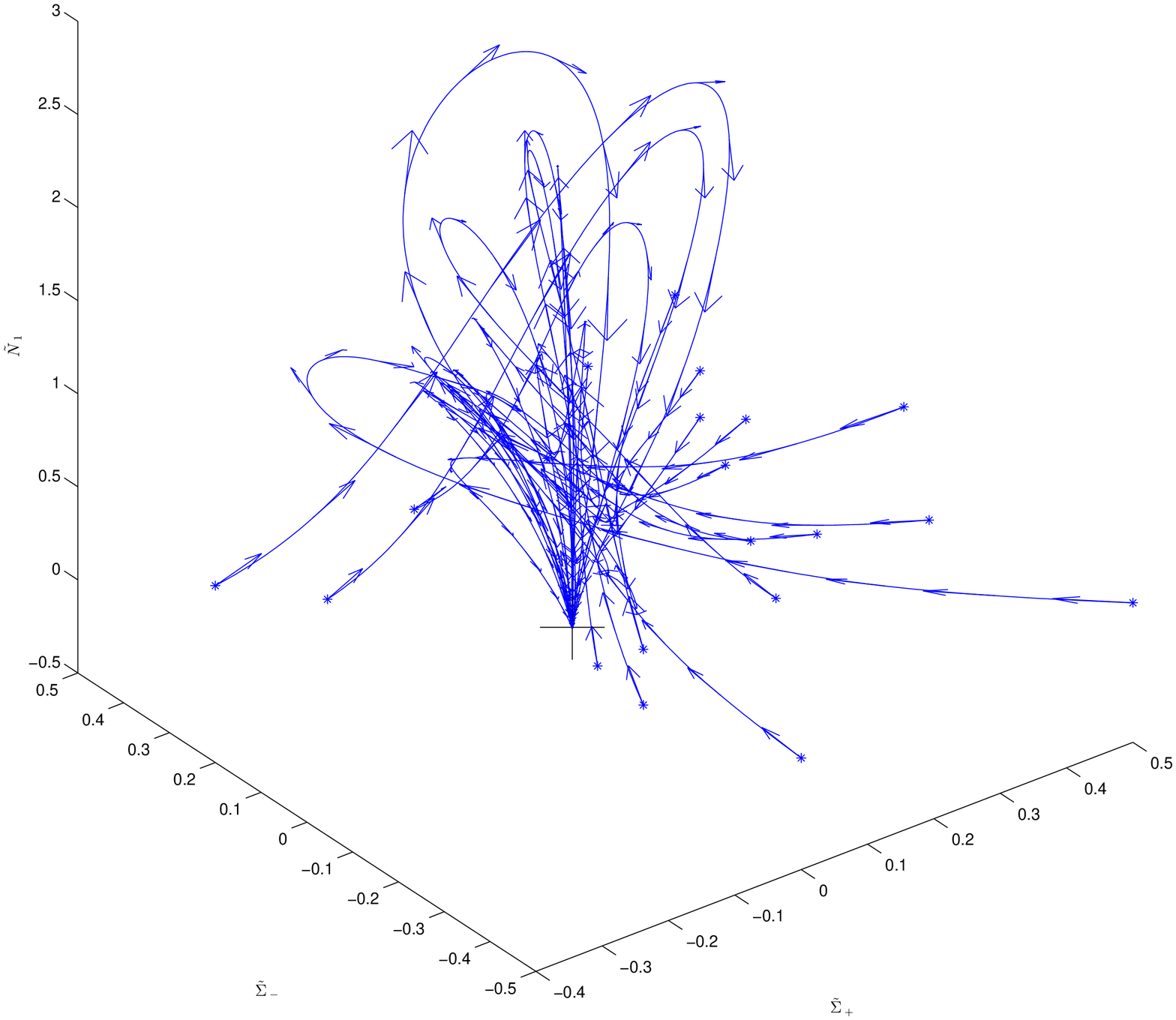}
\includegraphics*[scale = 0.40]{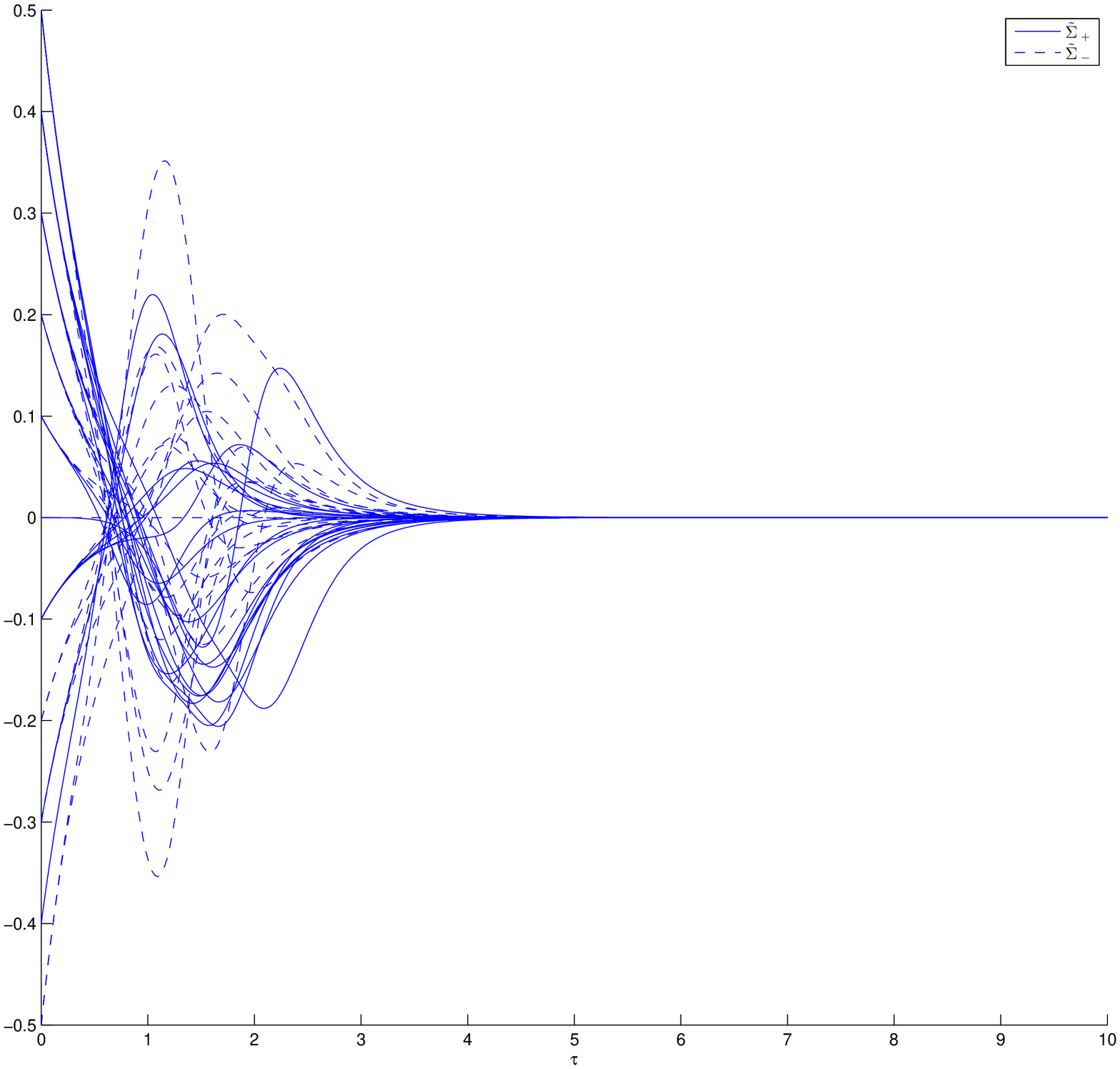}
\end{center}
\end{figure}

We display in Figs. \ref{fig1} and \ref{fig2} the results of a numerical experiment that show that $F_{\pm}$ are  local sinks of the system. 

\begin{figure}[H]
\begin{center}
\caption{This figure shows the dynamical system behavior for $\tilde{\xi}_{0} = 0$, $\tilde{\eta}_{0} = 1/2 $, and $w = 1/3$. The plus sign denotes the equilibrium point $P_{+}(II)$.  Numerical solutions were computed for $0 \leq \tau \leq 1000$. For clarity, we have displayed solutions for shorter timescales.} \label{fig3}
\includegraphics*[scale = 0.45]{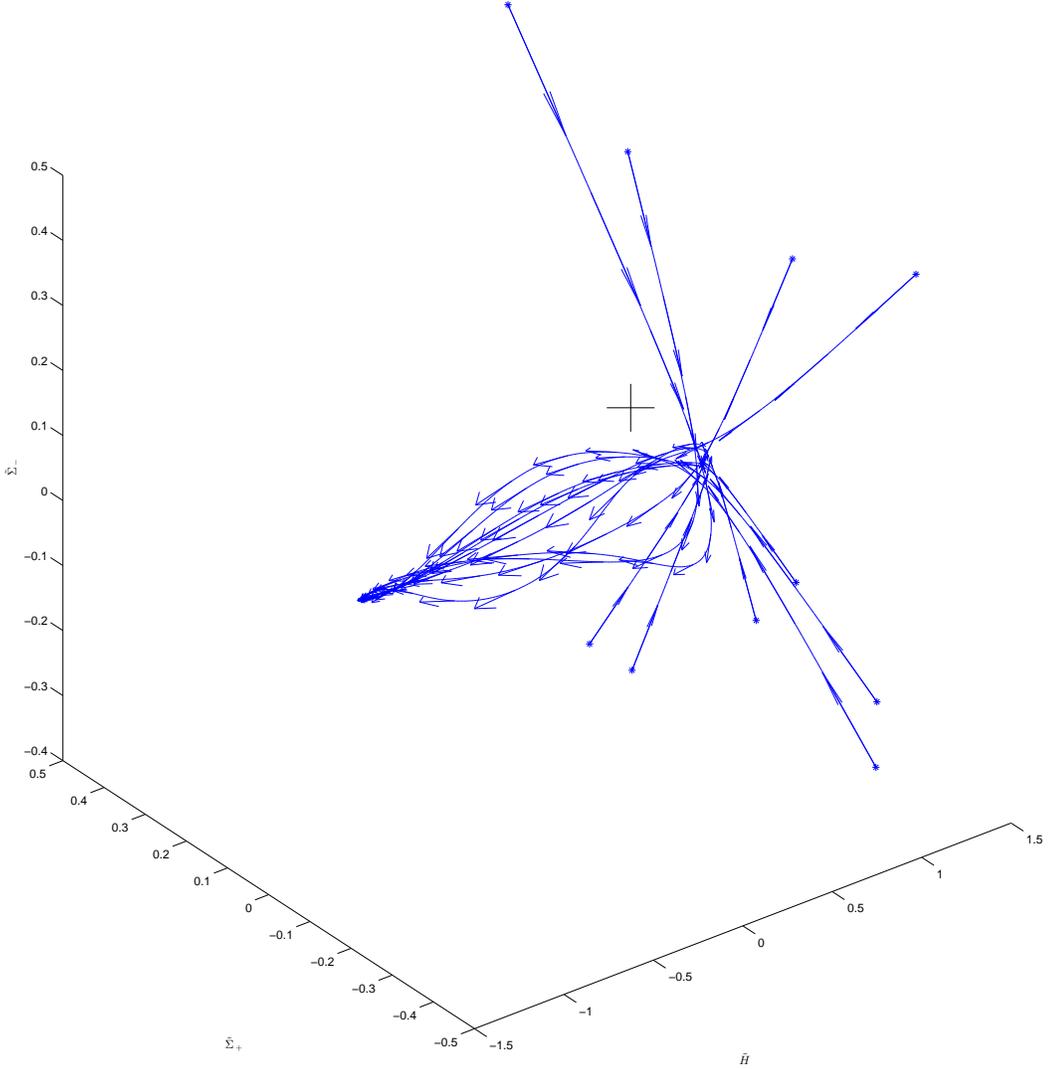}
\end{center}
\end{figure}

\begin{figure}[H]
\begin{center}
\caption{This figure shows the dynamical system behavior for $\tilde{\xi}_{0} = 2/9$, $\tilde{\eta}_{0} = 0 $, and $w = 1$. The plus sign denotes the equilibrium point $P_{+}(II)$.  Numerical solutions were computed for $0 \leq \tau \leq 1000$. For clarity, we have displayed solutions for shorter timescales.} \label{fig4}
\includegraphics*[scale = 0.45]{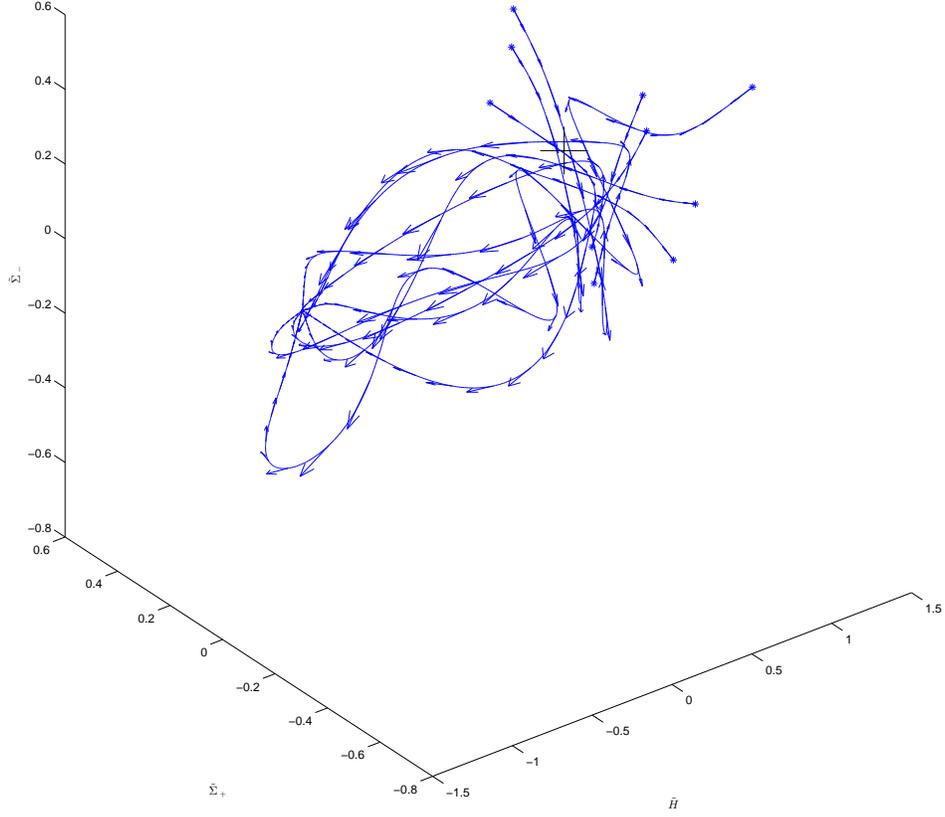}
\end{center}
\end{figure}

In Figs. \ref{fig3} and \ref{fig4}, we display the results of a numerical experiment that show that the points $P_{\pm}(II)$ correspond to saddles of the system.

\begin{figure}[H]
\begin{center}
\caption{This figure shows the dynamical system behavior for $\tilde{\xi}_{0} = 0$, $\tilde{\eta}_{0} = 0 $, and $w = 1$. The plus sign denotes the equilibrium point $\mathcal{J}_{+}$, while the diamond denotes the equilibrium point $\mathcal{J}_{-}$. Notice how all of the orbits are repelled by $\mathcal{J}_{+}$ but attracted by $\mathcal{J}_{-}$.  Numerical solutions were computed for $0 \leq \tau \leq 1000$. For clarity, we have displayed solutions for shorter timescales.} \label{fig5}
\includegraphics*[scale = 0.45]{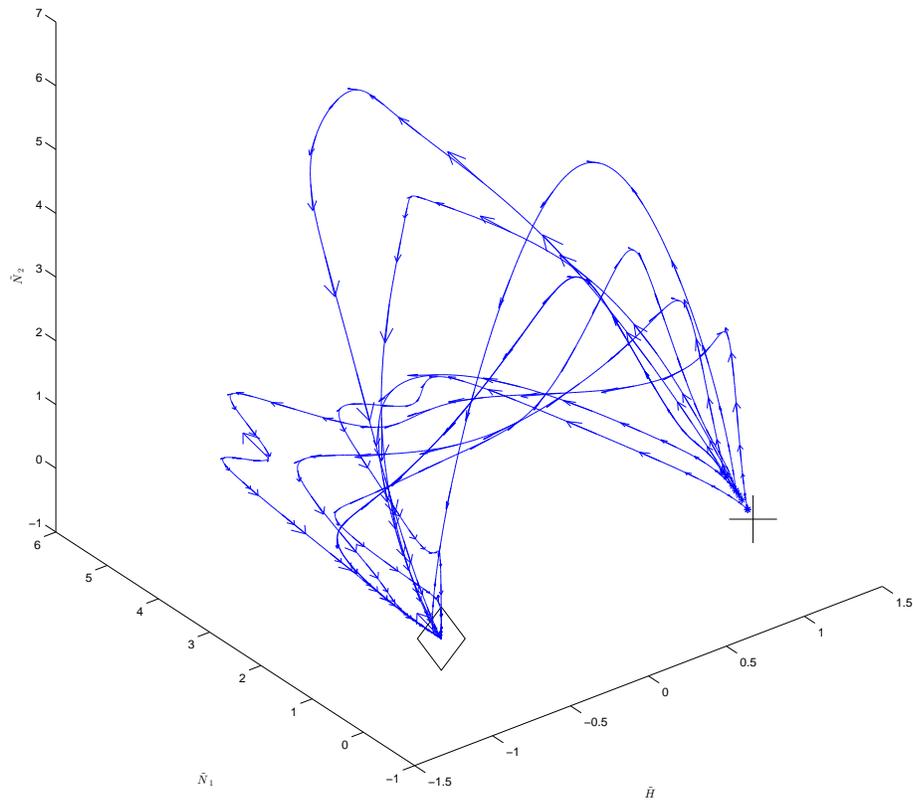}
\end{center}
\end{figure}

In Fig. \ref{fig5}, we display the results of a numerical experiment that show that the Jacobs disc set of equilibrium points $\mathcal{J}_{\pm}$ correspond to a local source and sink respectively. The same figure also shows the heteroclinic orbit behavior between $\mathcal{J}_{+}$ and $\mathcal{J}_{-}$.

\begin{figure}[H]
\begin{center}
\caption{This figure shows the dynamical system behavior for $\tilde{\xi}_{0} = 0$, $\tilde{\eta}_{0} = 0 $, and $w = 1/3$. The circular boundary defines the Kasner circle $\mathcal{K}_{+}$. In the last image, our numerical solutions for $\Delta$ as defined in Eq. \eqref{Deltadef} and $\tilde{\Omega}$ are displayed. Based on the conjecture discussed above, these results provide strong evidence that $M^{+}$ is indeed a past attractor for the dynamical systems.  Numerical solutions were computed for $0 \leq \tau \leq -1000$ For clarity, we have displayed solutions for shorter timescales. Note how in the first image half-ellipsoids form in the vertical direction as was predicted by the preceding analysis.}\label{fig6}
\includegraphics*[scale = 0.45]{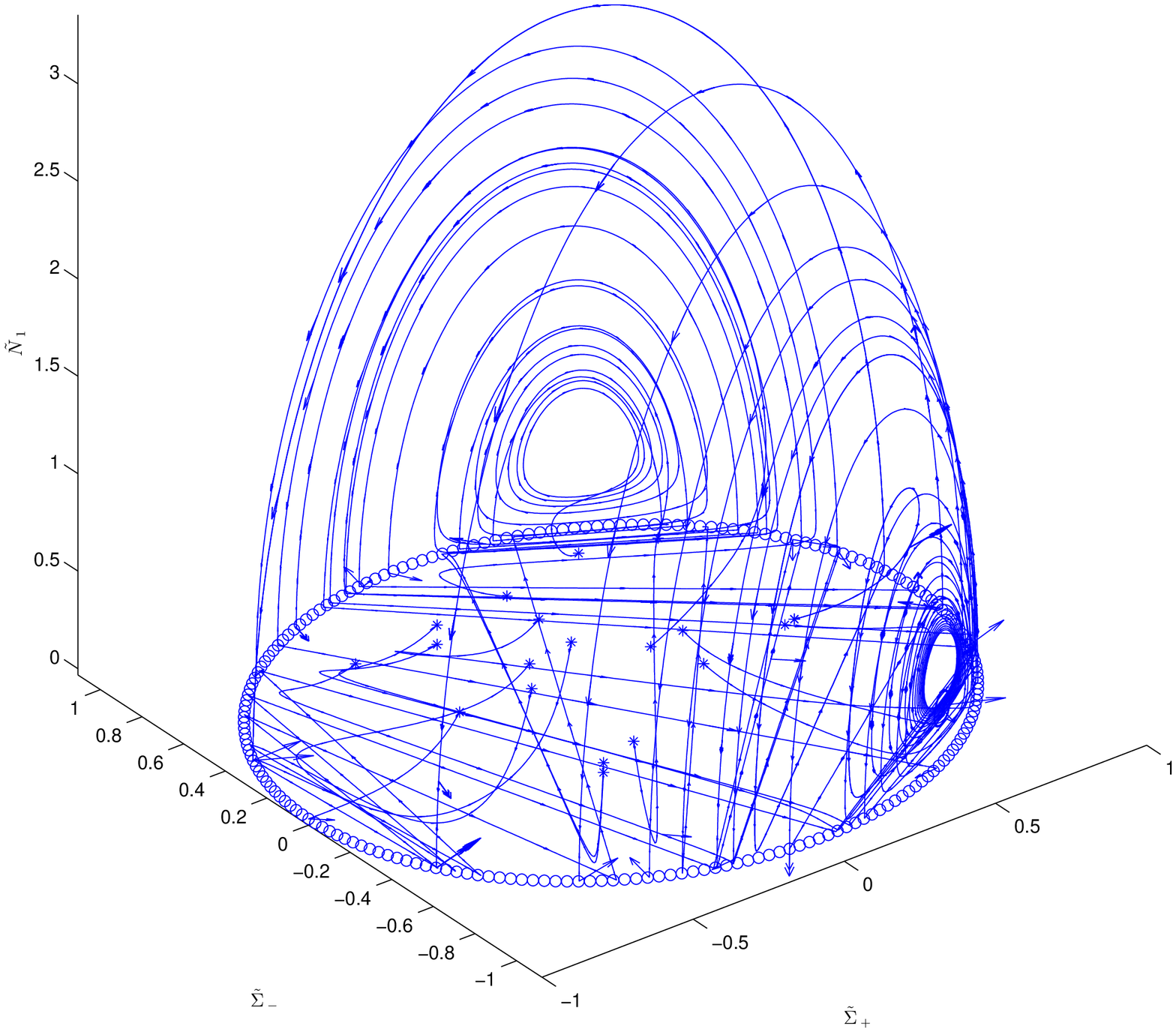}
\includegraphics*[scale = 0.45]{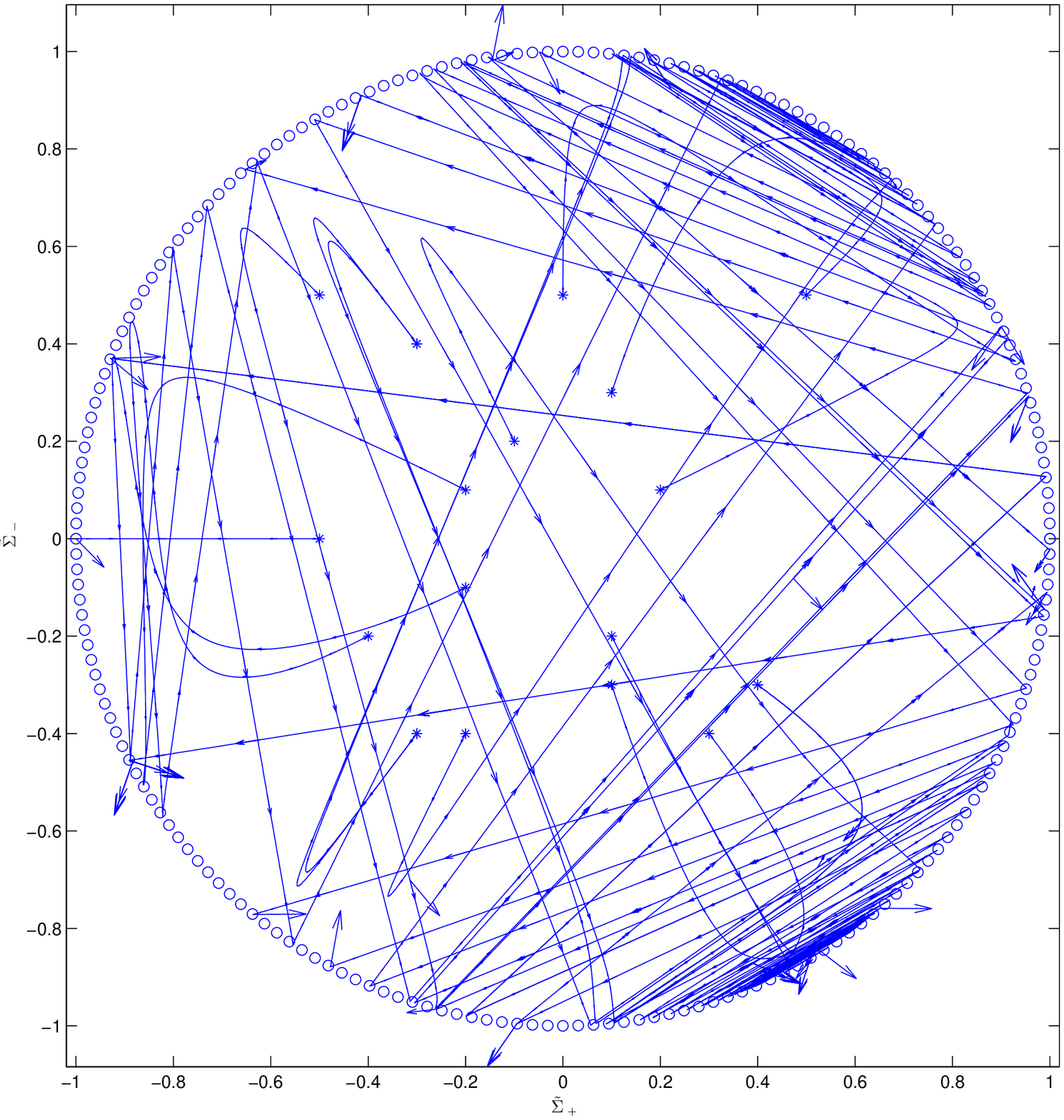}
\end{center}
\end{figure}

\begin{figure}[H]
\begin{center}
\caption{This figure shows the dynamical system behavior for $\tilde{\xi}_{0} = 0$, $\tilde{\eta}_{0} = 0 $, and $w = 1/3$. Our numerical solutions for $\Delta$ as defined in Eq. \eqref{Deltadef} and $\tilde{\Omega}$ are displayed. Based on the conjecture discussed above, these results provide strong evidence that $M^{+}$ is indeed a past attractor for the dynamical systems.  Numerical solutions were computed for $0 \leq \tau \leq -1000$ For clarity, we have displayed solutions for shorter timescales.}\label{figBIXexp}
\includegraphics*[scale = 0.45]{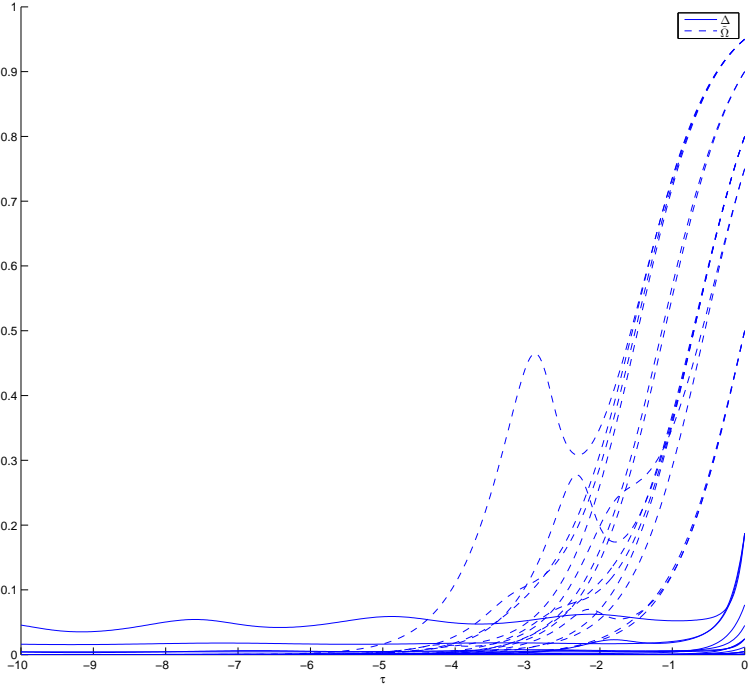}
\end{center}
\end{figure}

In Figs. \ref{fig6} and \ref{figBIXexp} we display the results of a numerical experiment that show the Mixmaster oscillatory behavior as the past orbits approach $\mathcal{K}_{+}$. 

\begin{figure}[H]
\begin{center}
\caption{This figure shows the dynamical system behavior for $\tilde{\xi}_{0} = 0$, $\tilde{\eta}_{0} = 0 $, and $w = 1/3$. In particular, it displays the heteroclinic orbits joining $\mathcal{K}_{+}$ to $\mathcal{K}_{-}$, where $\mathcal{K}_{+}$ is located at $\tilde{H} = 1$, and $\mathcal{K}_{-}$ is located at $\tilde{H} = -1$ in the figure. Numerical solutions were computed for $-1000 \leq \tau \leq 1000$. For clarity, we have displayed solutions for shorter timescales. }\label{fig7}
\includegraphics*[scale = 0.45]{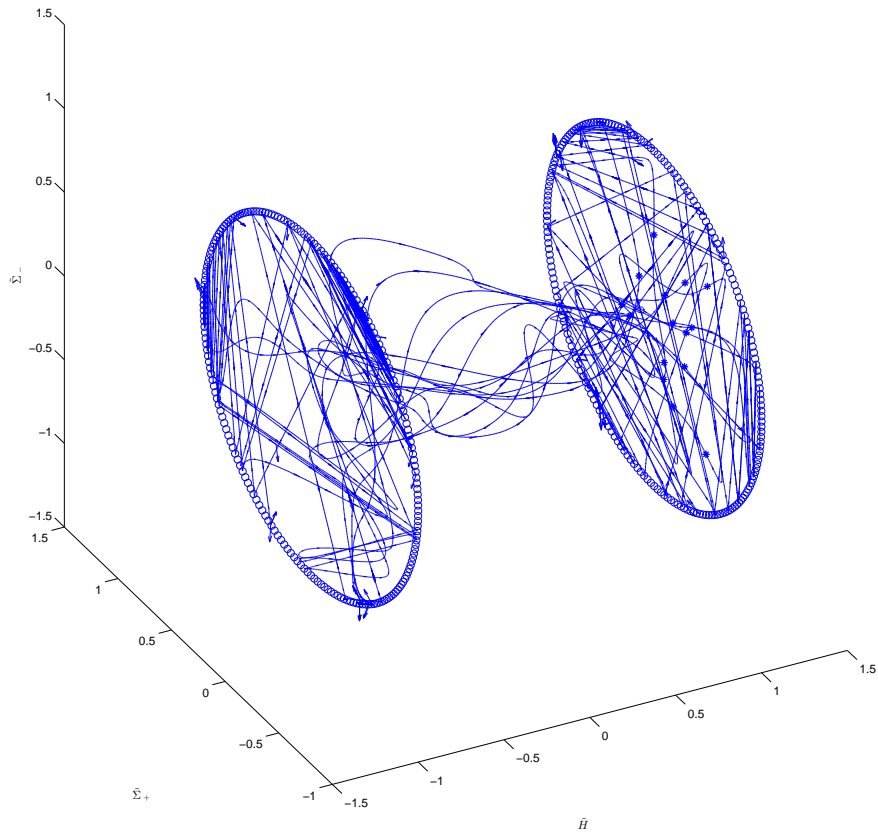}
\end{center}
\end{figure}

In Fig. \ref{fig7} we display the results of a numerical experiment that shows the heteroclinic orbits between $\mathcal{K}_{+}$ and $\mathcal{K}_{-}$. 


\section{Conclusions}
We have presented in this paper a comprehensive analysis of the dynamical behavior of a Bianchi Type IX viscous cosmology. We began by completing a detailed fix-point analysis which gave the local sinks, sources and saddles of the dynamical system. We then proceeded to analyze the global dynamics by finding the $\alpha$- and $\omega$-limit sets which gave an idea of the past and future asymptotic behavior of the system. The fixed points found were a flat FLRW solution, Bianchi Type $II$ solution, Kasner circle, Jacobs disc, Bianchi Type $VII_{0}$ solutions, and several closed FLRW solutions in addition to the Einstein static universe solution. Each equilibrium point was described in both its expanding and contracting epochs.

With respect to past asymptotic states, we were able to conclude that the Jacobs disc in the expanding epoch was a source of the system along with the flat FLRW solution in a contracting epoch. With respect to future asymptotic states, we were able to show that the flat FLRW solution in an expanding epoch along with the Jacobs disc in the contracting epoch were sinks of the system. We were also able to demonstrate a new result with respect to the Einstein static universe. Namely, we gave certain conditions on the parameter space such that the Einstein static universe has a stable subspace. We were however, not able to conclusively say anything about whether a closed FLRW model could be a past or future asymptotic state of the model.

The flat FLRW solution is clearly of primary importance with respect to modeling the present-day universe, which is observed to be very close to flat. We gave conditions in the parameter space for which this solution represents a saddle and a sink. When it is a saddle, the equilibrium point attracts along its stable manifold and repels along its unstable manifold. Therefore, some orbits will have an initial attraction to this point, but will eventually be repelled by it. In the case when it was found to be a sink, all orbits approach the equilibrium point in the future. Therefore, there exists a time period and two separate configurations for which our cosmological model will isotropize and be compatible with present-day observations of a high degree of isotropy in the cosmic microwave background.

\section{Acknowledgements}
One of us (ISK) would like to thank John D. Barrow for very helpful comments and suggestions.
This research was partially supported by a grant given to MCH from the Natural Sciences and Engineering Research Council of Canada.

\newpage 
\bibliography{sources}

\begin{thebibliography}{62}
\expandafter\ifx\csname natexlab\endcsname\relax\def\natexlab#1{#1}\fi
\expandafter\ifx\csname bibnamefont\endcsname\relax
  \def\bibnamefont#1{#1}\fi
\expandafter\ifx\csname bibfnamefont\endcsname\relax
  \def\bibfnamefont#1{#1}\fi
\expandafter\ifx\csname citenamefont\endcsname\relax
  \def\citenamefont#1{#1}\fi
\expandafter\ifx\csname url\endcsname\relax
  \def\url#1{\texttt{#1}}\fi
\expandafter\ifx\csname urlprefix\endcsname\relax\def\urlprefix{URL }\fi
\providecommand{\bibinfo}[2]{#2}
\providecommand{\eprint}[2][]{\url{#2}}

\bibitem[{\citenamefont{Wainwright and Ellis}(1997)}]{ellis}
\bibinfo{author}{\bibfnamefont{J.}~\bibnamefont{Wainwright}} \bibnamefont{and}
  \bibinfo{author}{\bibfnamefont{G.}~\bibnamefont{Ellis}},
  \emph{\bibinfo{title}{Dynamical Systems in Cosmology}}
  (\bibinfo{publisher}{Cambridge University Press}, \bibinfo{year}{1997}),
  \bibinfo{edition}{1st} ed.

\bibitem[{\citenamefont{Belinskii and Khalatnikov}(1970)}]{belinksii1}
\bibinfo{author}{\bibfnamefont{V.}~\bibnamefont{Belinskii}} \bibnamefont{and}
  \bibinfo{author}{\bibfnamefont{I.}~\bibnamefont{Khalatnikov}},
  \bibinfo{journal}{Soviet Physics JETP} \textbf{\bibinfo{volume}{30}},
  \bibinfo{pages}{1174} (\bibinfo{year}{1970}).

\bibitem[{\citenamefont{Belinskii et~al.}(1971)\citenamefont{Belinskii,
  Khalatnikov, and Lifshitz}}]{lifshitz}
\bibinfo{author}{\bibfnamefont{V.}~\bibnamefont{Belinskii}},
  \bibinfo{author}{\bibfnamefont{I.}~\bibnamefont{Khalatnikov}},
  \bibnamefont{and} \bibinfo{author}{\bibfnamefont{E.}~\bibnamefont{Lifshitz}},
  \bibinfo{journal}{Soviet Physics Uspekhi} \textbf{\bibinfo{volume}{13}},
  \bibinfo{pages}{745} (\bibinfo{year}{1971}).

\bibitem[{\citenamefont{Misner}(1969{\natexlab{a}})}]{misnerb}
\bibinfo{author}{\bibfnamefont{C.~W.} \bibnamefont{Misner}},
  \bibinfo{journal}{Physical Review} \textbf{\bibinfo{volume}{186}},
  \bibinfo{pages}{1319} (\bibinfo{year}{1969}{\natexlab{a}}).

\bibitem[{\citenamefont{Misner}(1969{\natexlab{b}})}]{misner}
\bibinfo{author}{\bibfnamefont{C.~W.} \bibnamefont{Misner}},
  \bibinfo{journal}{Physical Review Letters} \textbf{\bibinfo{volume}{22}},
  \bibinfo{pages}{1071} (\bibinfo{year}{1969}{\natexlab{b}}).

\bibitem[{\citenamefont{Matzner et~al.}(1970)\citenamefont{Matzner, Shepley,
  and Warren}}]{matzner1}
\bibinfo{author}{\bibfnamefont{R.}~\bibnamefont{Matzner}},
  \bibinfo{author}{\bibfnamefont{L.}~\bibnamefont{Shepley}}, \bibnamefont{and}
  \bibinfo{author}{\bibfnamefont{J.}~\bibnamefont{Warren}},
  \bibinfo{journal}{Annals of Physics} \textbf{\bibinfo{volume}{57}},
  \bibinfo{pages}{401} (\bibinfo{year}{1970}).

\bibitem[{\citenamefont{Ryan.}(1971)}]{ryan1}
\bibinfo{author}{\bibfnamefont{M.}~\bibnamefont{Ryan.}},
  \bibinfo{journal}{Annals of Physics} \textbf{\bibinfo{volume}{65}},
  \bibinfo{pages}{506} (\bibinfo{year}{1971}).

\bibitem[{\citenamefont{Ryan}(1971)}]{ryan2}
\bibinfo{author}{\bibfnamefont{M.}~\bibnamefont{Ryan}},
  \bibinfo{journal}{Annals of Physics} \textbf{\bibinfo{volume}{68}},
  \bibinfo{pages}{541} (\bibinfo{year}{1971}).

\bibitem[{\citenamefont{Ryan}(1972)}]{ryan3}
\bibinfo{author}{\bibfnamefont{M.}~\bibnamefont{Ryan}},
  \bibinfo{journal}{Annals of Physics} \textbf{\bibinfo{volume}{70}},
  \bibinfo{pages}{301} (\bibinfo{year}{1972}).

\bibitem[{\citenamefont{Arnowitt et~al.}(2008)\citenamefont{Arnowitt, Deser,
  and Misner}}]{admnew}
\bibinfo{author}{\bibfnamefont{R.}~\bibnamefont{Arnowitt}},
  \bibinfo{author}{\bibfnamefont{S.}~\bibnamefont{Deser}}, \bibnamefont{and}
  \bibinfo{author}{\bibfnamefont{C.~W.} \bibnamefont{Misner}},
  \bibinfo{journal}{General Relativity and Gravitation}
  \textbf{\bibinfo{volume}{40}}, \bibinfo{pages}{1997} (\bibinfo{year}{2008}).

\bibitem[{\citenamefont{Barrow and Matzner}(1980)}]{barrowmatz}
\bibinfo{author}{\bibfnamefont{J.}~\bibnamefont{Barrow}} \bibnamefont{and}
  \bibinfo{author}{\bibfnamefont{R.}~\bibnamefont{Matzner}},
  \bibinfo{journal}{Physical Review D} \textbf{\bibinfo{volume}{21}},
  \bibinfo{pages}{336} (\bibinfo{year}{1980}).

\bibitem[{\citenamefont{Barrow and Tipler}(1985)}]{barrowtip1}
\bibinfo{author}{\bibfnamefont{J.~D.} \bibnamefont{Barrow}} \bibnamefont{and}
  \bibinfo{author}{\bibfnamefont{F.~J.} \bibnamefont{Tipler}},
  \bibinfo{journal}{Monthly Notices of the Royal Astronomical Society}
  \textbf{\bibinfo{volume}{216}}, \bibinfo{pages}{395} (\bibinfo{year}{1985}).

\bibitem[{\citenamefont{Barrow et~al.}(1986)\citenamefont{Barrow, Galloway, and
  Tipler}}]{barrowtip2}
\bibinfo{author}{\bibfnamefont{J.~D.} \bibnamefont{Barrow}},
  \bibinfo{author}{\bibfnamefont{G.~J.} \bibnamefont{Galloway}},
  \bibnamefont{and} \bibinfo{author}{\bibfnamefont{F.~J.}
  \bibnamefont{Tipler}}, \bibinfo{journal}{Monthly Notices of the Royal
  Astronomical Society} \textbf{\bibinfo{volume}{223}}, \bibinfo{pages}{835}
  (\bibinfo{year}{1986}).

\bibitem[{\citenamefont{Barrow}(1988{\natexlab{a}})}]{barrownuc1}
\bibinfo{author}{\bibfnamefont{J.~D.} \bibnamefont{Barrow}},
  \bibinfo{journal}{Nuclear Physics B} \textbf{\bibinfo{volume}{296}},
  \bibinfo{pages}{697} (\bibinfo{year}{1988}{\natexlab{a}}).

\bibitem[{\citenamefont{Calogero and Heinzle}(2010)}]{caloheinzle}
\bibinfo{author}{\bibfnamefont{S.}~\bibnamefont{Calogero}} \bibnamefont{and}
  \bibinfo{author}{\bibfnamefont{J.}~\bibnamefont{Heinzle}},
  \bibinfo{journal}{Physical Review D} \textbf{\bibinfo{volume}{81}},
  \bibinfo{pages}{023520} (\bibinfo{year}{2010}).

\bibitem[{\citenamefont{Lin and Wald}(1989)}]{linwald}
\bibinfo{author}{\bibfnamefont{X.-F.} \bibnamefont{Lin}} \bibnamefont{and}
  \bibinfo{author}{\bibfnamefont{R.}~\bibnamefont{Wald}},
  \bibinfo{journal}{Physical Review D} \textbf{\bibinfo{volume}{40}},
  \bibinfo{pages}{3280} (\bibinfo{year}{1989}).

\bibitem[{\citenamefont{Lin and Wald}(1990)}]{linwald2}
\bibinfo{author}{\bibfnamefont{X.-F.} \bibnamefont{Lin}} \bibnamefont{and}
  \bibinfo{author}{\bibfnamefont{R.}~\bibnamefont{Wald}},
  \bibinfo{journal}{Physical Review D} \textbf{\bibinfo{volume}{41}},
  \bibinfo{pages}{2444} (\bibinfo{year}{1990}).

\bibitem[{\citenamefont{Wald}(1983)}]{waldix}
\bibinfo{author}{\bibfnamefont{R.}~\bibnamefont{Wald}},
  \bibinfo{journal}{Physical Review D} \textbf{\bibinfo{volume}{28}},
  \bibinfo{pages}{2118} (\bibinfo{year}{1983}).

\bibitem[{\citenamefont{Burd et~al.}(1990)\citenamefont{Burd, Buric, and
  Ellis}}]{burdburicellis}
\bibinfo{author}{\bibfnamefont{A.}~\bibnamefont{Burd}},
  \bibinfo{author}{\bibfnamefont{N.}~\bibnamefont{Buric}}, \bibnamefont{and}
  \bibinfo{author}{\bibfnamefont{G.}~\bibnamefont{Ellis}},
  \bibinfo{journal}{General Relativity and Gravitation}
  \textbf{\bibinfo{volume}{22}}, \bibinfo{pages}{349} (\bibinfo{year}{1990}).

\bibitem[{\citenamefont{Rugh and Jones}(1990)}]{rughjones}
\bibinfo{author}{\bibfnamefont{S.}~\bibnamefont{Rugh}} \bibnamefont{and}
  \bibinfo{author}{\bibfnamefont{B.}~\bibnamefont{Jones}},
  \bibinfo{journal}{Physics Letters A} \textbf{\bibinfo{volume}{147}},
  \bibinfo{pages}{353} (\bibinfo{year}{1990}).

\bibitem[{\citenamefont{Uggla and Zur-Muhlen}(1990)}]{ugglazur}
\bibinfo{author}{\bibfnamefont{C.}~\bibnamefont{Uggla}} \bibnamefont{and}
  \bibinfo{author}{\bibfnamefont{H.}~\bibnamefont{Zur-Muhlen}},
  \bibinfo{journal}{Classical and Quantum Gravity}
  \textbf{\bibinfo{volume}{7}}, \bibinfo{pages}{1365} (\bibinfo{year}{1990}).

\bibitem[{\citenamefont{Cornish and Levin}(1997{\natexlab{a}})}]{cornishlevin}
\bibinfo{author}{\bibfnamefont{N.}~\bibnamefont{Cornish}} \bibnamefont{and}
  \bibinfo{author}{\bibfnamefont{J.}~\bibnamefont{Levin}},
  \bibinfo{journal}{Physical Review Letters} \textbf{\bibinfo{volume}{78}},
  \bibinfo{pages}{998} (\bibinfo{year}{1997}{\natexlab{a}}).

\bibitem[{\citenamefont{Cornish and Levin}(1997{\natexlab{b}})}]{cornishlevin2}
\bibinfo{author}{\bibfnamefont{N.}~\bibnamefont{Cornish}} \bibnamefont{and}
  \bibinfo{author}{\bibfnamefont{J.}~\bibnamefont{Levin}},
  \bibinfo{journal}{Physical Review D} \textbf{\bibinfo{volume}{55}},
  \bibinfo{pages}{7489} (\bibinfo{year}{1997}{\natexlab{b}}).

\bibitem[{\citenamefont{Rendall}(1994)}]{rendall}
\bibinfo{author}{\bibfnamefont{A.}~\bibnamefont{Rendall}},
  \bibinfo{journal}{Annals of Physics} \textbf{\bibinfo{volume}{233}},
  \bibinfo{pages}{82} (\bibinfo{year}{1994}).

\bibitem[{\citenamefont{Rendall}(1997)}]{rendall2}
\bibinfo{author}{\bibfnamefont{A.}~\bibnamefont{Rendall}},
  \bibinfo{journal}{Classical and Quantum Gravity}
  \textbf{\bibinfo{volume}{14}}, \bibinfo{pages}{2341} (\bibinfo{year}{1997}).

\bibitem[{\citenamefont{Van Den~Hoogen and Olasagasti}(1999)}]{vandenolas}
\bibinfo{author}{\bibfnamefont{R.}~\bibnamefont{Van Den~Hoogen}}
  \bibnamefont{and}
  \bibinfo{author}{\bibfnamefont{I.}~\bibnamefont{Olasagasti}},
  \bibinfo{journal}{Physical Review D} \textbf{\bibinfo{volume}{59}},
  \bibinfo{pages}{1} (\bibinfo{year}{1999}).

\bibitem[{\citenamefont{Ringstr\"{o}m}(2001)}]{ringstrom1}
\bibinfo{author}{\bibfnamefont{H.}~\bibnamefont{Ringstr\"{o}m}},
  \bibinfo{journal}{Annales Henri Poincare} \textbf{\bibinfo{volume}{2}},
  \bibinfo{pages}{405} (\bibinfo{year}{2001}).

\bibitem[{\citenamefont{De~Oliveira et~al.}(2002)\citenamefont{De~Oliveira,
  Ozorio~de Almeida, Dami\~{o}~Soares, and Tonini}}]{deoliv1}
\bibinfo{author}{\bibfnamefont{H.}~\bibnamefont{De~Oliveira}},
  \bibinfo{author}{\bibfnamefont{A.}~\bibnamefont{Ozorio~de Almeida}},
  \bibinfo{author}{\bibfnamefont{I.}~\bibnamefont{Dami\~{o}~Soares}},
  \bibnamefont{and} \bibinfo{author}{\bibfnamefont{E.}~\bibnamefont{Tonini}},
  \bibinfo{journal}{Physical Review D} \textbf{\bibinfo{volume}{65}},
  \bibinfo{pages}{835111} (\bibinfo{year}{2002}).

\bibitem[{\citenamefont{Barrow et~al.}(2003)\citenamefont{Barrow, Ellis,
  Maartens, and Tsagas}}]{barrowellismaartenstsagas}
\bibinfo{author}{\bibfnamefont{J.}~\bibnamefont{Barrow}},
  \bibinfo{author}{\bibfnamefont{G.}~\bibnamefont{Ellis}},
  \bibinfo{author}{\bibfnamefont{R.}~\bibnamefont{Maartens}}, \bibnamefont{and}
  \bibinfo{author}{\bibfnamefont{C.}~\bibnamefont{Tsagas}},
  \bibinfo{journal}{Classical and Quantum Gravity}
  \textbf{\bibinfo{volume}{20}}, \bibinfo{pages}{155} (\bibinfo{year}{2003}).

\bibitem[{\citenamefont{Heinzle et~al.}(2005)\citenamefont{Heinzle, R{\"o}hr,
  and Uggla}}]{heinzleuggla}
\bibinfo{author}{\bibfnamefont{J.}~\bibnamefont{Heinzle}},
  \bibinfo{author}{\bibfnamefont{N.}~\bibnamefont{R{\"o}hr}}, \bibnamefont{and}
  \bibinfo{author}{\bibfnamefont{C.}~\bibnamefont{Uggla}},
  \bibinfo{journal}{Physical Review D} \textbf{\bibinfo{volume}{71}},
  \bibinfo{pages}{1} (\bibinfo{year}{2005}).

\bibitem[{\citenamefont{Heinzle and Uggla}(2009{\natexlab{a}})}]{heinzleuggla2}
\bibinfo{author}{\bibfnamefont{J.}~\bibnamefont{Heinzle}} \bibnamefont{and}
  \bibinfo{author}{\bibfnamefont{C.}~\bibnamefont{Uggla}},
  \bibinfo{journal}{Classical and Quantum Gravity}
  \textbf{\bibinfo{volume}{26}}, \bibinfo{pages}{075016}
  (\bibinfo{year}{2009}{\natexlab{a}}).

\bibitem[{\citenamefont{Heinzle and Uggla}(2009{\natexlab{b}})}]{heinzleuggla3}
\bibinfo{author}{\bibfnamefont{J.}~\bibnamefont{Heinzle}} \bibnamefont{and}
  \bibinfo{author}{\bibfnamefont{C.}~\bibnamefont{Uggla}},
  \bibinfo{journal}{Classical and Quantum Gravity}
  \textbf{\bibinfo{volume}{26}}, \bibinfo{pages}{075015}
  (\bibinfo{year}{2009}{\natexlab{b}}).

\bibitem[{\citenamefont{Calogero and Heinzle}(2011)}]{caloheinzle2}
\bibinfo{author}{\bibfnamefont{S.}~\bibnamefont{Calogero}} \bibnamefont{and}
  \bibinfo{author}{\bibfnamefont{J.}~\bibnamefont{Heinzle}},
  \bibinfo{journal}{Physica D: Nonlinear Phenomena}
  \textbf{\bibinfo{volume}{240}}, \bibinfo{pages}{636} (\bibinfo{year}{2011}).

\bibitem[{\citenamefont{Barrow and Yamamoto}(2012)}]{barrowyama}
\bibinfo{author}{\bibfnamefont{J.}~\bibnamefont{Barrow}} \bibnamefont{and}
  \bibinfo{author}{\bibfnamefont{K.}~\bibnamefont{Yamamoto}},
  \bibinfo{journal}{Physical Review D} \textbf{\bibinfo{volume}{85}},
  \bibinfo{pages}{083505} (\bibinfo{year}{2012}).

\bibitem[{\citenamefont{Uggla}(2013)}]{ugglanew}
\bibinfo{author}{\bibfnamefont{C.}~\bibnamefont{Uggla}},
  \bibinfo{journal}{International Journal of Modern Physics D}
  \textbf{\bibinfo{volume}{22}}, \bibinfo{pages}{1330002}
  (\bibinfo{year}{2013}).

\bibitem[{\citenamefont{Gr{\o}n and Hervik}(2007)}]{hervik}
\bibinfo{author}{\bibfnamefont{{\O}.}~\bibnamefont{Gr{\o}n}} \bibnamefont{and}
  \bibinfo{author}{\bibfnamefont{S.}~\bibnamefont{Hervik}},
  \emph{\bibinfo{title}{Einstein's General Theory of Relativity: With Modern
  Applications in Cosmology}} (\bibinfo{publisher}{Springer},
  \bibinfo{year}{2007}), \bibinfo{edition}{1st} ed.

\bibitem[{\citenamefont{Parnovskii}(1977)}]{parnovskii}
\bibinfo{author}{\bibfnamefont{S.}~\bibnamefont{Parnovskii}},
  \bibinfo{journal}{Journal of Experimental and Theoretical Physics}
  \textbf{\bibinfo{volume}{45}}, \bibinfo{pages}{423} (\bibinfo{year}{1977}).

\bibitem[{\citenamefont{Misner}(1968)}]{misnervisc1}
\bibinfo{author}{\bibfnamefont{C.~W.} \bibnamefont{Misner}},
  \bibinfo{journal}{Astrophysical Journal} \textbf{\bibinfo{volume}{151}},
  \bibinfo{pages}{431} (\bibinfo{year}{1968}).

\bibitem[{\citenamefont{Misner}(1967)}]{misnervisc2}
\bibinfo{author}{\bibfnamefont{C.~W.} \bibnamefont{Misner}},
  \bibinfo{journal}{Physical Review Letters} \textbf{\bibinfo{volume}{19}},
  \bibinfo{pages}{533} (\bibinfo{year}{1967}).

\bibitem[{\citenamefont{Caderni and Fabbri}(1978)}]{cadfab}
\bibinfo{author}{\bibfnamefont{N.}~\bibnamefont{Caderni}} \bibnamefont{and}
  \bibinfo{author}{\bibfnamefont{R.}~\bibnamefont{Fabbri}},
  \bibinfo{journal}{Physics Letters A} \textbf{\bibinfo{volume}{67}},
  \bibinfo{pages}{19} (\bibinfo{year}{1978}).

\bibitem[{\citenamefont{Caderni and Fabbri}(1979)}]{cadfab2}
\bibinfo{author}{\bibfnamefont{N.}~\bibnamefont{Caderni}} \bibnamefont{and}
  \bibinfo{author}{\bibfnamefont{R.}~\bibnamefont{Fabbri}},
  \bibinfo{journal}{Physical Review D} \textbf{\bibinfo{volume}{20}},
  \bibinfo{pages}{1251} (\bibinfo{year}{1979}).

\bibitem[{\citenamefont{Banerjee and Santos}(1984)}]{banerjeesantos}
\bibinfo{author}{\bibfnamefont{A.}~\bibnamefont{Banerjee}} \bibnamefont{and}
  \bibinfo{author}{\bibfnamefont{N.}~\bibnamefont{Santos}},
  \bibinfo{journal}{General Relativity and Gravitation}
  \textbf{\bibinfo{volume}{16}}, \bibinfo{pages}{217} (\bibinfo{year}{1984}).

\bibitem[{\citenamefont{Banerjee et~al.}(1990)\citenamefont{Banerjee, Sanyal,
  and Chakraborty}}]{banerjee2}
\bibinfo{author}{\bibfnamefont{A.}~\bibnamefont{Banerjee}},
  \bibinfo{author}{\bibfnamefont{A.}~\bibnamefont{Sanyal}}, \bibnamefont{and}
  \bibinfo{author}{\bibfnamefont{S.}~\bibnamefont{Chakraborty}},
  \bibinfo{journal}{Astrophysics and Space Science}
  \textbf{\bibinfo{volume}{166}}, \bibinfo{pages}{259} (\bibinfo{year}{1990}).

\bibitem[{\citenamefont{Chakraborty and Chakraborty}(2001)}]{chakraborty}
\bibinfo{author}{\bibfnamefont{N.}~\bibnamefont{Chakraborty}} \bibnamefont{and}
  \bibinfo{author}{\bibfnamefont{S.}~\bibnamefont{Chakraborty}},
  \bibinfo{journal}{Nuovo Cimento della Societa Italiana di Fisica B}
  \textbf{\bibinfo{volume}{116}}, \bibinfo{pages}{191} (\bibinfo{year}{2001}).

\bibitem[{\citenamefont{Pradhan et~al.}(2005)\citenamefont{Pradhan, Srivastav,
  and Yadav}}]{prasriya}
\bibinfo{author}{\bibfnamefont{A.}~\bibnamefont{Pradhan}},
  \bibinfo{author}{\bibfnamefont{S.}~\bibnamefont{Srivastav}},
  \bibnamefont{and} \bibinfo{author}{\bibfnamefont{M.}~\bibnamefont{Yadav}},
  \bibinfo{journal}{Astrophysics and Space Science}
  \textbf{\bibinfo{volume}{298}}, \bibinfo{pages}{419} (\bibinfo{year}{2005}).

\bibitem[{\citenamefont{Bali and Yadav}(2005)}]{baliyadav}
\bibinfo{author}{\bibfnamefont{R.}~\bibnamefont{Bali}} \bibnamefont{and}
  \bibinfo{author}{\bibfnamefont{M.}~\bibnamefont{Yadav}},
  \bibinfo{journal}{Pramana-Journal of Physics} \textbf{\bibinfo{volume}{64}},
  \bibinfo{pages}{187} (\bibinfo{year}{2005}).

\bibitem[{\citenamefont{Ellis and MacCallum}(1969)}]{ellismac}
\bibinfo{author}{\bibfnamefont{G.}~\bibnamefont{Ellis}} \bibnamefont{and}
  \bibinfo{author}{\bibfnamefont{M.}~\bibnamefont{MacCallum}},
  \bibinfo{journal}{Comm. Math. Phys} \textbf{\bibinfo{volume}{12}},
  \bibinfo{pages}{108} (\bibinfo{year}{1969}).

\bibitem[{\citenamefont{van~den Hoogen and Coley}(1995)}]{vdh}
\bibinfo{author}{\bibfnamefont{R.}~\bibnamefont{van~den Hoogen}}
  \bibnamefont{and} \bibinfo{author}{\bibfnamefont{A.}~\bibnamefont{Coley}},
  \bibinfo{journal}{Classical and Quantum Gravity}
  \textbf{\bibinfo{volume}{12}}, \bibinfo{pages}{2335} (\bibinfo{year}{1995}).

\bibitem[{\citenamefont{Kohli and Haslam}(2013{\natexlab{a}})}]{isk1}
\bibinfo{author}{\bibfnamefont{I.~S.} \bibnamefont{Kohli}} \bibnamefont{and}
  \bibinfo{author}{\bibfnamefont{M.~C.} \bibnamefont{Haslam}},
  \bibinfo{journal}{Phys. Rev. D} \textbf{\bibinfo{volume}{87}},
  \bibinfo{pages}{063006} (\bibinfo{year}{2013}{\natexlab{a}}),
  \urlprefix\url{http://link.aps.org/doi/10.1103/PhysRevD.87.063006}.

\bibitem[{\citenamefont{Kohli and Haslam}(2013{\natexlab{b}})}]{isk2}
\bibinfo{author}{\bibfnamefont{I.~S.} \bibnamefont{Kohli}} \bibnamefont{and}
  \bibinfo{author}{\bibfnamefont{M.~C.} \bibnamefont{Haslam}},
  \bibinfo{journal}{Phys. Rev. D} \textbf{\bibinfo{volume}{88}},
  \bibinfo{pages}{063518} (\bibinfo{year}{2013}{\natexlab{b}}),
  \urlprefix\url{http://link.aps.org/doi/10.1103/PhysRevD.88.063518}.

\bibitem[{\citenamefont{Barrow}(1988{\natexlab{b}})}]{barrownuc2}
\bibinfo{author}{\bibfnamefont{J.~D.} \bibnamefont{Barrow}},
  \bibinfo{journal}{Nuclear Physics B} \textbf{\bibinfo{volume}{310}},
  \bibinfo{pages}{743} (\bibinfo{year}{1988}{\natexlab{b}}).

\bibitem[{\citenamefont{Barrow}(1982)}]{barrownuc3}
\bibinfo{author}{\bibfnamefont{J.~D.} \bibnamefont{Barrow}},
  \bibinfo{journal}{Monthly Notices of the Royal Astronomical Society}
  \textbf{\bibinfo{volume}{199}}, \bibinfo{pages}{45} (\bibinfo{year}{1982}).

\bibitem[{\citenamefont{Hewitt et~al.}(2001)\citenamefont{Hewitt, Bridson, and
  Wainwright}}]{hewittbridsonwainwright}
\bibinfo{author}{\bibfnamefont{C.}~\bibnamefont{Hewitt}},
  \bibinfo{author}{\bibfnamefont{R.}~\bibnamefont{Bridson}}, \bibnamefont{and}
  \bibinfo{author}{\bibfnamefont{J.}~\bibnamefont{Wainwright}},
  \bibinfo{journal}{General Relativity and Gravitation}
  \textbf{\bibinfo{volume}{33}}, \bibinfo{pages}{65} (\bibinfo{year}{2001}).

\bibitem[{\citenamefont{Anosov et~al.}(1997)\citenamefont{Anosov, Aranson,
  Arnold, Bronshtein, Grines, and Il'yashenko}}]{arnolddyn}
\bibinfo{author}{\bibfnamefont{D.}~\bibnamefont{Anosov}},
  \bibinfo{author}{\bibfnamefont{S.~K.} \bibnamefont{Aranson}},
  \bibinfo{author}{\bibfnamefont{V.}~\bibnamefont{Arnold}},
  \bibinfo{author}{\bibfnamefont{I.}~\bibnamefont{Bronshtein}},
  \bibinfo{author}{\bibfnamefont{V.}~\bibnamefont{Grines}}, \bibnamefont{and}
  \bibinfo{author}{\bibfnamefont{Y.}~\bibnamefont{Il'yashenko}},
  \emph{\bibinfo{title}{Ordinary Differential Equations and Smooth Dynamical
  Systems}} (\bibinfo{publisher}{Springer-Verlag}, \bibinfo{year}{1997}),
  \bibinfo{edition}{3rd} ed.

\bibitem[{\citenamefont{Wainwright and Hsu}(1989)}]{whpaper}
\bibinfo{author}{\bibfnamefont{J.}~\bibnamefont{Wainwright}} \bibnamefont{and}
  \bibinfo{author}{\bibfnamefont{L.}~\bibnamefont{Hsu}},
  \bibinfo{journal}{Classical and Quantum Gravity}
  \textbf{\bibinfo{volume}{6}}, \bibinfo{pages}{1409} (\bibinfo{year}{1989}).

\bibitem[{\citenamefont{Ellis et~al.}(2012)\citenamefont{Ellis, Maartens, and
  MacCallum}}]{elliscosmo}
\bibinfo{author}{\bibfnamefont{G.~F.} \bibnamefont{Ellis}},
  \bibinfo{author}{\bibfnamefont{R.}~\bibnamefont{Maartens}}, \bibnamefont{and}
  \bibinfo{author}{\bibfnamefont{M.~A.} \bibnamefont{MacCallum}},
  \emph{\bibinfo{title}{Relativistic Cosmology}} (\bibinfo{publisher}{Cambridge
  University Press}, \bibinfo{year}{2012}), \bibinfo{edition}{1st} ed.

\bibitem[{\citenamefont{Einstein}(1952)}]{einstein}
\bibinfo{author}{\bibfnamefont{A.}~\bibnamefont{Einstein}},
  \emph{\bibinfo{title}{The Principle of Relativity}}
  (\bibinfo{publisher}{Dover}, \bibinfo{year}{1952}), \bibinfo{edition}{1st}
  ed.

\bibitem[{\citenamefont{Lema\^{i}tre}(2013)}]{lemaitre1}
\bibinfo{author}{\bibfnamefont{G.}~\bibnamefont{Lema\^{i}tre}},
  \bibinfo{journal}{General Relativity and Gravitation}
  \textbf{\bibinfo{volume}{45}}, \bibinfo{pages}{1635} (\bibinfo{year}{2013}).

\bibitem[{\citenamefont{Lema\^{i}tre}(1931)}]{lemaitre2}
\bibinfo{author}{\bibfnamefont{G.}~\bibnamefont{Lema\^{i}tre}},
  \bibinfo{journal}{Monthly Notices of the Royal Astronomical Society}
  \textbf{\bibinfo{volume}{91}}, \bibinfo{pages}{490} (\bibinfo{year}{1931}).

\bibitem[{\citenamefont{Eddington}(1930)}]{eddington}
\bibinfo{author}{\bibfnamefont{A.}~\bibnamefont{Eddington}},
  \bibinfo{journal}{Monthly Notices of the Royal Astronomical Society}
  \textbf{\bibinfo{volume}{90}}, \bibinfo{pages}{668} (\bibinfo{year}{1930}).

\bibitem[{\citenamefont{Hewitt and Wainwright}(1993)}]{hewwain}
\bibinfo{author}{\bibfnamefont{C.}~\bibnamefont{Hewitt}} \bibnamefont{and}
  \bibinfo{author}{\bibfnamefont{J.}~\bibnamefont{Wainwright}},
  \bibinfo{journal}{Classical and Quantum Gravity}
  \textbf{\bibinfo{volume}{10}}, \bibinfo{pages}{99} (\bibinfo{year}{1993}).

\bibitem[{\citenamefont{LeBlanc et~al.}(1995)\citenamefont{LeBlanc, Kerr, and
  Wainwright}}]{leblanc3}
\bibinfo{author}{\bibfnamefont{V.}~\bibnamefont{LeBlanc}},
  \bibinfo{author}{\bibfnamefont{D.}~\bibnamefont{Kerr}}, \bibnamefont{and}
  \bibinfo{author}{\bibfnamefont{J.}~\bibnamefont{Wainwright}},
  \bibinfo{journal}{Classical and Quantum Gravity}
  \textbf{\bibinfo{volume}{12}}, \bibinfo{pages}{513} (\bibinfo{year}{1995}).

\end{thebibliography}
\end{document}